\documentclass[vecphys]{svmult}

\usepackage{makeidx}         
\usepackage{graphicx}        
\usepackage{multicol}        
\usepackage[bottom]{footmisc}

\usepackage{amsmath}            
\usepackage{color}			

\makeindex             

\newcommand{\myA} {{\cal A}}
\newcommand{\myD} {{\cal D}}
\newcommand{\myH} {{\cal H}}
\newcommand{\myJ} {{\cal J}}
\newcommand{\myL} {{\cal L}}

\newcommand{\myO} {{\cal O}}
\newcommand{\myP} {{\cal P}}
\newcommand{\myQ} {{\cal Q}}

\newcommand{\mysigma} {{\sigma^{(1)}}}
\newcommand{\mysigmai} {{\sigma_i^{(1)}}}
\newcommand{\mysigmaj} {{\sigma_j^{(1)}}}

\newcommand{\mytau} {{\sigma^{(2)}}}
\newcommand{\mytaui} {{\sigma_i^{(2)}}}
\newcommand{\mytauj} {{\sigma_j^{(2)}}}

\newcommand{\sigman}  {{\sigma^{(n)}}}

\newcommand{\qtilda}  {{\tilde{q}}}

\newcommand{\Med}[1]{\left\langle #1 \right\rangle}

\begin{document}

\title*{
Some aspects of infinite range models of spin glasses: theory and
numerical simulations
}
\author{Alain Billoire\inst{1} }
\titlerunning{Infinite range models of spin glasses}
\institute{Service de physique th\'eorique,
CEA Saclay,
91191 Gif-sur-Yvette FRANCE
\texttt{billoire@spht.saclay.cea.fr
}
}

%
%
\maketitle

\date{}
The expression ``spin glasses" was originally coined to describe
metallic alloys of a non-magnetic metal with few, randomly
substituted, magnetic impurities. Experimental evidences where
obtained for a low temperature ``spin-glass" phase characterized by a
non-periodic freezing of the magnetic moments (the spins) with a very
slow, and strongly history dependent, response to external
perturbations (this later aspect lead more recently to many
fascinating developments). The theoretical analysis of this phenomenon
lead to the celebrated Edwards--Anderson model~\cite{EdAn} of
spin-glasses: classical spins on the sites of a regular lattice with
random interactions between nearest neighbor spins. However, after
more than thirty years of intense studies, the very nature of the low
temperature phase of the Edwards--Anderson model in three dimensions
is still debated, even in the simple case of Ising spins. Two main
competing theories exist: the mean field approach originating from the
work of Sherrington and Kirkpatrick~\cite{ShKi}, and the so-called
``droplet"~\cite{Droplets} or scaling theory of spin glasses.

The mean field approach is the application to this problem of the
conventional approach to phase transitions in statistical physics: one
first builds a mean field theory after identifying the proper order
parameter, solve it (usually a straightforward task) and then study
the fluctuations around the mean field solution. Usually, fluctuations
turn out to have mild effects for space dimensions above the so-called
upper critical dimension (up to infinite space dimension, where mean
field is exact). Below the upper critical dimension fluctuations have
major effects and non-perturbative techniques are needed to handle
them. The second item of this agenda (solving the mean field
equations) led, with spin glasses, to severe unexpected difficulties,
and revealed a variety of new fascinating phenomena. The last step is
the subject of the so called ``replica field theory'', which is still
facing formidable difficulties.

These notes are an introduction to the physics of the infinite range
version of the Edwards--Anderson model, the so-called
Sherrington--Kirkpatrick model, namely a model of classical spins that
are not embedded in Euclidean space, with all pairs of spins
interacting with a random interaction. If there is no more debate
whether Parisi famous solution of the Sherrington--Kirkpatrick model
in the infinite volume limit is correct, much less is known, as
mentioned before, about the Edwards--Anderson model in three
dimensions, with numerical simulations as one of our main sources of
knowledge. It is accordingly important to test the various methods of
analysis proposed for the Edwards--Anderson model, in the
Sherrington--Kirkpatrick model case first.

In a first part, I motivate and introduce the Edwards--Anderson and
Sherrington--Kirkpatrick models.  In the second part, I sketch the
analytical solution of the Sherrington--Kirkpatrick model, following
Parisi\cite{Pa1,Pa2,Pa3,Pa4}. I~next give the physical interpretation
of this solution.  This is a vast subject, and I concentrate on the
major points and give references for more developments.  The third
part presents the numerical simulation approach and compares some
numerical results to theoretical expectations.  The last part, more
detailed, is about the specific problem of finite size effects for the
free energy, which is interesting for both theoretical and practical
point of views.  I have left aside several very interesting aspects,
like the problem of chaos~\cite{RizCr,BiMa} the TAP
approach~\cite{TAP} (see~\cite{TAPnum} for numerical results) and the
computation of the complexity~\cite{Complexity}.

There are many books and review articles about spin glasses and
related phenomena: One may start with the text book by Fischer and
Hertz~\cite{FiHe}, the review by Binder and Young~\cite{BiYo}, which
gives a very complete account of the situation in 1986 both
experimental, theoretical and numerical with many detailed analytical
computations, and (at a higher level) the book by M\'ezard, Parisi,
and Virasoro~\cite{MePaVi}.  More recent references
include~\cite{CaCa} and~\cite{Sh}.  The recent book by de Dominicis
and Giardina~\cite{DeDoGi} gives a very compete exposition of the
replica field theory.  Reviews on various aspects of the physics of
spin glasses can be found
in~\cite{KaRi,MaPaRuLo,MaPaRiTeRULo}. The non
equilibrium behavior of spin glasses have been the subject of intense
work during recent years, see~\cite{BoCuKuMe,Cu,CrRit}
for reviews.

\section{Introduction}
\label{sec:1}

The Edwards--Anderson Ising (EAI) model is the classical Ising model with
quenched random interactions.\index{Edwards--Anderson model}
The  Hamiltonian is

\begin{equation}
{\myH}=-\sum_{<i,j>}  J_{i,j} \sigma_i \sigma_j - H \sum_i \sigma_i\;,
\label{Hamiltonian}
\end{equation}
where the variables $\sigma_i=\pm 1$ are Ising spins living on the
sites of a regular square lattice in $d$ dimensions.  $H$ is the
magnetic field.  The interaction involves all nearest neighbor pairs
of spins, with a strength $J_{i,j}$ that depends on the particular
link $<i,j>$. The $J_{i,j}$'s are quenched variables, namely they do not
fluctuate. They have been drawn (once for ever) independently from a
unique probability distribution $P(J)$ with mean $J_0$ and square deviation
$J^2$.  In what follows, I will consider the case $J_0=0$ and choose
$J^2=1$. The later choice is just the choice of the unit of
temperature.

For a fixed set of $J_{i,j}$'s, denoted by $\myJ$, one can compute
thermodynamic average values, for example the average internal energy
$E_J=<E>$, where the symbol $<\cdots>$ denotes the thermodynamics
average (with given $\myJ$). In general the results depend on $\myJ$,
and should be averaged over the $\myJ$ distribution. I note by
$\overline{\cdots}$ this disorder average.

This model was proposed~\cite{EdAn} as a simple model that captures the
essential of the physics of spin glasses, and in particular of
magnetic spin glasses. Those are alloys with magnetic impurities
randomly distributed inside a non magnetic matrix.  The interaction
between the impurities is the RKKY interaction that oscillates with
the distance.  Since the positions of the impurities are random, their
interactions are random too and one is led to the Edwards--Anderson
model (see e.g.~\cite{EdAn,FiHe,BiYo} for more than this abrupt
summary).  If this very simple model is to explain convincingly the
behavior of real spin glasses, its physics should obviously not depend
(too much) on the specific disorder distribution used, as soon as it
has zero mean and unit square deviation.  This disorder distribution
is usually Gaussian (this leads to simpler analytical computations) or
binary with values $\pm 1$ (this leads to faster computer programs,
using the multi-spin coding technique).

The same model can be generalized to vector spins with two or three
components.  Those are called the XY and Heisenberg Edwards Anderson
models respectively. Most real spin glasses are indeed Heisenberg spin
glass with anisotropic interactions (due to the lattice structure, the
interaction is not the rotational invariant
$\vec{\sigma_i}\vec{\sigma_j}$).

In the zero magnetic field case ($H=0$), there is now an agreement
(based mostly on numerical simulations) between spin glass physicists
that the EAI model has no transition at finite $T$ in two dimensions,
and a transition for three dimensions and above~\footnote{The value of
the lower critical dimension, between 2 and 3, is not yet settled.}.  In the
low temperature phase the spins are frozen, with $<\sigma_i>\neq 0$,
but with a random pattern, and $\sum_i <\sigma_i>$ is of order
$1/\sqrt{N}$, where $N$ is the number of spins. There is accordingly
no spontaneous magnetization. There is no hidden magnetization either:
the spins follow no periodic pattern (like e.g.~in an
anti-ferromagnet), and it is furthermore very likely that the spin
orientations are completely reshuffled as soon as one varies the
temperature~\footnote{This is the so called temperature chaos effect.}.
This is summarized by the statement: The EAI model is not a disguised
ferromagnet.

The statement $<\sigma_i>\neq 0$ needs to be made more precise. For a
finite system, it means that $<\sigma_i>$ is nonzero when observed over
a time scale $t<<t_{erg}(N)$, where $t_{erg}(N)$ is called the ergodic
time.  At some later  time $t$, the system will eventually tunnel from the
current equilibrium state, to a state where   $<\sigma_i>$ is
reversed. However the dynamics
of the EAI model is extremely slow and $t_{erg}(N)$ is enormous in the
low $T$ phase, as soon as one has few hundred spins, to the point that
with any real spin glass, this time is much larger than the duration
of any experiment.

In absence of magnetization, we have to find another order parameter
for  the glass order, one uses the so called Edwards--Anderson
parameter, namely

\begin{equation}
q_{EA}=
\lim_{t\to\infty}\lim_{N\to\infty} \frac{1}{N}
\sum_i <\sigma_i(t_0) \sigma_i(t+t_0)>\;,
\label{qEA}
\end{equation}
that is nonzero below the spin glass transition and zero above.  On a
finite system, $C(t)=1/N \sum_i <\sigma_i(t_0) \sigma_i(t+t_0)>$, will
rapidly decrease from the starting value $C(1)=1$, have a plateau of
height $q_{EA}$ for a long time (of order $t_{erg}$) before
dropping. As $N$ grows, the plateau becomes longer and longer.  The so
defined $q_{EA}$ turns out to depend weakly on $\myJ$ for large
systems. This is a so called self averaging quantity (see
Subsec~\ref{subsec:11}).

One may also consider two independent copies of the system, $\{\mysigmai\}$ and
$\{\mytaui\}$, with the same disorder instance $\myJ$ (such copies are
called real replica, or sometimes clones, in order to distinguish them
from the replica of the replica method, see
Subsec.~\ref{subsec:10}), and consider the probability distribution

\begin{equation}
P_{\myJ}(q)= <\delta(q-\frac{1}{N}\sum_i \mysigmai \mytaui)>\;,
\label{overlap}
\end{equation}
namely the probability distribution of the overlap $q= 1/N \sum_i
\mysigmai \mytaui$ between the two systems. A nonzero $q_{EA}$
corresponds to two peaks in $P_{\myJ}(q)$ centered at $q=\pm q_{EA}$.
We will see later that, at least for the SK model, $P_{\myJ}(q)$ has
more structure than this double peak.

Clearly,  
$q_{EA}=<M>^2$ for a ferromagnet in $d$
dimensions (still at zero magnetic field), where $<M>$ is the
spontaneous magnetization. On a finite system, $P(q)$ is
made of two Gaussian centered around $\pm <M>^2$. Between the two
peaks, $P(q)$ is exponentially small~\cite{BeBiJa}, of order $\exp(-2\myA L^{d-1})$,
where $\myA$ is the interface tension, and $L$ the linear
dimension of the system.

In ferromagnets, the transition happens when the magnetization $M$
acquires a nonzero expectation value, and accordingly the
susceptibility $\chi=N <M^2>$ becomes infinite as $T$ decreases
towards $T_c$. In spin glasses the spin susceptibility stays finite as
$T \to T_c$, but the spin-glass susceptibility $\chi_{SG} = N
\overline{<q^2>}$ diverges. There is however a difference: in Ising
ferromagnets, one can define a connected susceptibility~\footnote{In
this formula $<M>$ is to be interpreted as an average restricted over
a time interval of length $<<t_{erg}(N)$, or simply as $<|M|>$.}
$N(<M^2>-<M>^2)$ that is finite in the low $T$ phase. For Ising spin
glasses the analog connected susceptibility diverges, as $N\to \infty$
in the whole low $T$ phase~\footnote{\samepage{It is usually defined
as $\sum_{i,j}
\overline{(<\sigma_i\sigma_j>-<\sigma_i><\sigma_j>)^2}$, sometimes
simply as $N(\overline{<q^2>}-\overline{<q>}^2)$.   .}}

\section{The Sherrington--Kirkpatrick model}

\index{Sherrington--Kirkpatrick model}
This model~\cite{ShKi} is a simplified version of the Edwards--Anderson model,
where all spins are directly coupled. The Hamiltonian is

\begin{equation}
{\cal H}=-\sum_{1\leq i<j \leq N} \frac{J_{i,j}}{\sqrt{N}} \sigma_i
\sigma_j - H \sum_i \sigma_i\;,
\end{equation}
where $N$ is the number of spins. The $J_{i,j}$'s are again drawn from
a unique probability distribution $P(\myJ)$ with zero mean and unit
square deviation. The factor $1/\sqrt{N}$ will ensure a finite limit
for the internal energy per spin $e_N(T)$ as $N\to\infty$.  It is in
some sense a model in infinite dimension since a given spin is coupled
to $N$ spins, with $N\to\infty$ in the thermodynamic limit. The model
has ``infinite connectivity''. The infinite connectivity will ensure
that the mean field method gives the exact result.  As we will show,
the SK model can be solved exactly in the thermodynamic limit, and the
result is independent of the choice made for the disorder distribution
$P(\myJ)$.

The standard method to solve this model is the famous replica
trick~\footnote{There is an alternative method called the cavity
method, see~\cite{MePaVi,MePaVi86}.}.\index{Replica trick} One is
interested in computing the average free energy $-\overline {\beta
F_{\myJ}}=\overline {\ln Z_{\myJ}}$, where $\beta=1/T$ is the inverse
temperature, in units such that $k_B=1$.  This is done by considering
the partition function of $n$ identical un-coupled copies of the
system, with the same instance of the disorder $\myJ$. Such copies are
called replica. The partition function is simply the $n$th power of
$Z_{\myJ}$, $Z_{\myJ}^n$. Continuing this function defined for integer
$n\geq 1$ to a function of the real variable $n$, one has, at least
for finite $N$,

\begin{eqnarray}
-\beta F= \overline{\ln Z_{\myJ}} =\lim _{n\to 0} {\frac{\overline
{Z_{\myJ}^n}-1}{n}}\;.
\end{eqnarray}

$Z_J^n$ can be written as

\begin{eqnarray}
Z_J^n &=&Tr_{\sigma}^{[n]} \exp(-\beta (\myH^{[n]}_{\myJ}))\\
\nonumber
&=&Tr_{\sigma}^{[n]}
\exp(-\beta (\myH_{\myJ}(\mysigma)+\cdots+\myH_{\myJ}(\sigman)))\;,
\label{replica1}
\end{eqnarray}
where $\sigma^{(1)}$ represents the $N$ spins of the first replica
(namely $\{\sigma^{(1)}_i\}$), $\sigma^{(2)}$ the $N$ spins of the
second replica, \ldots, and $Tr_{\sigma}^{[n]}$is the trace over the
$n N$ spin variables.  The average over the $n(n-1)/2$ $J_{i,j}$
variables are independent, and the disorder average factorizes as a
product of terms of the form

\begin{eqnarray}
\overline{
\exp\Biggl(\frac{\beta J_{i,j}  X_{i,j}}{\sqrt{N}}\Biggr)}
=\exp\Biggl(\sum_{p=1}^{\infty}\frac{\beta^p X_{i,j}^p[J]_p}{p! N^{p/2}}\Biggr)\;,
\label{cumulant}
\end{eqnarray}
where

\begin{equation}
X_{i,j} =\sigma^{(1)}_i \sigma^{(1)}_j +\sigma^{(2)}_i \sigma^{(2)}_j
+\cdots +\sigma^{(n)}_i \sigma^{(n)}_j=
\sum_{a=1}^n \sigma^{(a)}_i \sigma^{(a)}_j\;, 
\end{equation}
and the $[J]_p$'s are the successive cumulants of the disorder distribution $P(\myJ)$,

\begin{eqnarray}
{[J]}_1 &=& \overline{J} \\
\nonumber
{[J]}_2 &=& \overline{(J-\overline{J})^2}\\
\nonumber
{[J]}_3 &=& \overline{(J-\overline{J})^3}\\
\nonumber
{[J]}_4 &=& \overline{(J-\overline{J})^4}-3{\overline{(J-\overline{J})^2}}^2\\
\nonumber
\dots
\end{eqnarray}

Since we have $\overline{J}=0$ and $\overline{J^2}=1$, one is led to the result

\begin{eqnarray}
\overline{
\exp\Biggl(\frac{\beta J_{i,j}  X_{i,j}}{\sqrt{N}}\Biggr)}
=\exp\Biggl(\frac{\beta^2 X_{i,j}^2}{2 N}+\cdots\Biggr)\;.
\label{average}
\end{eqnarray}

The neglected terms in the exponent are of order $1/N^2$ and will not
contribute to the thermodynamics limit~\footnote{They are altogether
absent if the distribution is Gaussian.}, and accordingly the physics
is independent of the disorder distribution.  One obtains for the
disorder averaged partition function:

\begin{eqnarray}
\overline {Z_{\myJ}^n}&=&Tr_{\sigma}^{[n]} \exp\Bigl(\frac{\beta^2}{2N} \sum_{i<j} \
(\sum_{b=1}^n \sigma^{(b)}_i \sigma^{(b)}_j )^2
+\beta H\sum_i \sum_{b=1}^n \sigma^{(b)}_i
\Bigr)\\
\nonumber
&=&Tr_{\sigma}^{[n]} \exp\Bigl(\frac{\beta^2}{2N}
\sum_{a<b}(\sum_i \sigma^{(a)}_i \sigma^{(b)}_i)^2
+\beta H\sum_i \sum_{b=1}^n \sigma^{(b)}_i\\
\nonumber
&+& (n N-n^2) \frac{\beta ^2}{4}\Bigr)\;.
\end{eqnarray}

The average over the disorder has been performed
analytically, but now the $n$ replicas are coupled. In order to proceed further,
one uses the formula

\begin{eqnarray}
\sqrt{\frac{N\beta^2}{2 \pi}} \int_{-\infty}^{+\infty} dq \ \
\exp (-\frac{N\beta^2q^2}{2})
\quad 
\exp(q\beta^2 X)&=&\exp(\frac{\beta^2X^2}{2N} )\;,
\label{magic}
\end{eqnarray}
introducing $n(n-1)/2$ auxiliary real
variables $q_{a,b}$ ($a<b$), with the result

\begin{eqnarray}
\overline {Z_{\myJ}^n}&=& \Bigl[\prod_{a<b} \sqrt{\frac{N\beta^2}{2\pi}} \int dq_{a,b}\Bigr]
\ Tr_{\sigma}^{[n]} 
\exp\Bigl(-\frac{N\beta^2}{2}\sum_{a<b} q_{a,b}^2
\\
\nonumber
&+&\beta^2 \sum_{a<b} q_{a,b}\sum_i \sigma^{(a)}_i
\sigma^{(b)}_i 
+\beta H\sum_i \sum_{b=1}^n \sigma^{(b)}_i
+ (n N-n^2) \frac{\beta ^2}{4}\Bigr)\;.
\label{replica2}
\end{eqnarray}

The variables $q_{a,b}$ have been defined for $a<b$.  In the
following, it will be sometimes convenient to define $q_{a,b}$ for $a\geq b$
also, as $q_{a,b}=q_{b,a}$ and $q_{a,a}=0$.  For a given replica index
$a$, the trace over the spins factorizes as

\begin{eqnarray}
& &\prod_{i=1}^N Tr_{\sigma_i^{(a)}}\  e^{\displaystyle\beta^2
\sum_{a<b}  q_{a,b}\sum_i \sigma^{(a)}_i \sigma^{(b)}_i
+\beta H\sum_i \sum_{b=1}^n \sigma^{(b)}_i}\\
\nonumber
&=&
\Biggl(Tr_{S^{(a)}}\ e^{\displaystyle \beta^2\sum_{a<b}   
q_{a,b} S^{(a)} S^{(b)}
+\beta H \sum_{b=1}^n S^{(b)}}
\Biggr)^N\;,
\end{eqnarray}
where $S^{(a)}$ is any of the $\sigma_i^{(a)}$'s, or alternatively the
spin of a single spin system, which is replicated $n$ times. Thus~\footnote{From
now on, we omit the $n^2$ terms in the exponent, since they do not
contribute in the $n\to 0$ limit.}

\begin{eqnarray}
\label{replica4}
\overline {Z_{\myJ}^n}&=& \Bigl[\prod_{a<b} \sqrt{\frac{N\beta^2}{2\pi}} \int dq_{a,b}\Bigr]
\exp\Biggl(-\frac{N\beta^2}{2}\sum_{a<b} q_{a,b}^2 \\
\nonumber
&+&N \log\Bigl(Tr^{[n]}_{S} [\exp(\beta^2 \sum_{a<b} q_{a,b} S^{(a)}S^{(b)}
+\beta H \sum_{b=1}^n S^{(b)}
)]\Bigr)
+ n N \frac{\beta ^2}{4}\Biggr)\;.
\end{eqnarray}
The formula has a 
particularly simple form, with all dependence in $N$
explicit

\begin{eqnarray}
\overline {Z_{\myJ}^n}&=& \Bigl[\prod_{a<b} \sqrt{\frac{N\beta^2}{2\pi}} \int dq_{a,b}\Bigr]\
\exp(-N \beta\myA (\{q_{a,b}\}))\;.
\label{replica11}
\end{eqnarray}

Up to now our derivation is exact (for a Gaussian disorder) in the
$n\to0$ limit.  We  now make the saddle point (or steepest descent)
approximation, which gives the correct $N\to\infty$ behavior.
Assuming the existence of a unique absolute maximum of the integrand at
location $\{q^{SP}_{a,b}\}$, one has

\begin{eqnarray}
\overline {Z_{\myJ}^n}&\approx& 
\exp(-N \beta \myA (\{q^{SP}_{a,b}\}))\;.
\end{eqnarray}

By assumption, all partial derivatives of $\myA$, $\partial
\myA/\partial q_{a,b}$, are zero at the saddle point, and the matrix
of the second derivatives, the Hessian, has only nonnegative
eigenvalues.  The free energy of the original SK model is thus simply
related to the saddle point of the $q_{a,b}$ integral representation
of the partition function of the $n$ times replicated model (with a
reckless interchange of limits),

\begin{equation}
f=\lim_{N\to\infty}\frac{F}{N}= \lim_{n\to 0} \frac{1}{n} \myA(\{q^{SP}_{a,b}\})\;.
\end{equation}

Note that the saddle point equations can be written as
self consistent equations, involving a single spin system

\begin{eqnarray}
q^{(SP)}_{a,b} &=&\frac{Tr_{S} [S^{(a)}S^{(b)}
\exp(\displaystyle\beta^2 \sum_{a<b}
q_{a,b}^{(SP)}
S^{(a)}S^{(b)} +\beta H \sum_{b=1}^n S^{(b)} )]} {Tr_{S}
[\exp(\displaystyle\beta^2 \sum_{a<b} q^{(SP)}_{a,b}S^{(a)}S^{(b)} +\beta H
\sum_{b=1}^n S^{(b)} )]}\;.
\end{eqnarray}

The replica method is not limited to the evaluation of the free
energy. It can be used to compute the average (thermodynamic average
and disorder average) of any function of the spins. Let $\myO(\sigma)$
be a function of the spins $\{\sigma_i\}$, we have by
definition, for any disorder sample:

\begin{equation}
<\myO(\sigma)>=\frac{
Tr_{\sigma}^{[1]} \myO(\sigma^{(1)}) \exp(-\beta (\myH^{[1]}_{\myJ}))}
{Z_{\myJ}}\;. 
\end{equation}

Multiplying both numerator and denominator by $Z^{n-1}$, and
letting~\footnote{We did add $n-1$ replica to the one real replica and
then let $n\to 0$.} $n \to 0$, we have

\begin{equation}
<\myO(\sigma)>=\lim_{n\to 0}
Tr_{\sigma}^{[n]} \myO(\sigma^{(1)}) \exp(-\beta (\myH^{[n]}_{\myJ}))\;.
\end{equation}

The right hand side is of a form whose disorder average is readily computed
using~(\ref{average}).  The method is extended readily to the disorder average
of products like $<\myO(\sigma)><\myP(\sigma)>$. Introducing two real
replica, we have indeed

\begin{equation}
<\myO(\sigma)>
<\myP(\sigma)>
=\frac{
Tr_{\sigma}^{[2]} \myO(\sigma^{(1)})  \myP(\sigma^{(2)}) 
\exp(-\beta (\myH^{[2]}_{\myJ}))}
{Z_{\myJ}^2}\;. 
\end{equation}

Multiplying both numerator and denominator by $Z^{n-2}$, and
letting $n \to 0$, we have

\begin{equation}
<\myO(\sigma)><\myP(\sigma)>
=\lim_{n\to 0}
Tr_{\sigma}^{[n]} \myO(\sigma^{(1)})  \myP(\sigma^{(2)}) 
\exp(-\beta (\myH^{[n]}_{\myJ}))\;.
\end{equation}
which is again of a form whose disorder average is readily computed
using~(\ref{average}).

\subsection{Interpretation of the  $q_{a,b}$ variables}

The $q_{a,b}$ variables have been introduced formally. Their value at
the saddle point has  a simple interpretation~\cite{Pa5} in terms of the
overlaps between real replicas as defined in~(\ref{overlap}). Consider
two clones $\sigma^{(1)}$ and $\sigma^{(1)}$ and compute the
generating function $G(y)$\footnote{This is a convenient way to
evaluate at once the expression $\overline{<(1/N\sum_i \sigma^{(1)}_i
\sigma^{(2)}_j)^k>}$ for all values of $k$.}.

\begin{equation}
G(y)=\overline{<\exp( \frac{y \beta^2}{N} \sum_i \sigma^{(1)}_i \sigma^{(2)}_i)>}\;.
\end{equation}

We have,  

\begin{equation}
{\Bigl<e^{\displaystyle \frac{y \beta^2}{N} \sum_i \sigma^{(1)}_i \sigma^{(2)}_i}\Bigr>}=
\frac{1}{Z^2}
Tr_{\sigma}^{(2)} e^{\displaystyle\frac{y \beta^2}{N} \sum_i \sigma^{(1)}_i
\sigma^{(2)}_i -\beta \myH_{\myJ}^{[2]}}\;.
\label{replica6}
\end{equation}

Multiplying the numerator and denominator by $Z^{n-2}$, and
letting $n\to 0$, one obtains after averaging over the disorder

\begin{eqnarray}
\label{replica10}
G(y)&=& Z^{-n}  \Bigl[\prod_{a<b} \sqrt{\frac{N\beta^2}{2\pi}} \int dq_{a,b}\Bigr]
\exp\Biggl(\displaystyle{y \beta^2} q_{1,2}
-\frac{N\beta^2}{2}\sum_{a<b} q_{a,b}^2 \\
\nonumber
&+&N \log\Bigl(Tr_{S}^{[n]} [\exp(\beta^2 \sum_{a<b} q_{a,b} S^{(a)}S^{(b)}
+\beta H \sum_{b=1}^n S^{(b)}
)]\Bigr)
+ n N \frac{\beta ^2}{4}\Biggr)\;,
\label{replica8}
\end{eqnarray}
with $Z^n=1$ since $n\to 0$.

At leading order in $1/N$, the saddle point is $y$ independent, and accordingly

\begin{equation}
\overline{<\exp( \frac{y\beta^2}{N} \sum_i \sigma^{(1)}_i \sigma^{(2)}_i)>}
=\exp(y\beta^2 q_{1,2}^{SP})\;.
\end{equation}

If the saddle point is not symmetric, namely if the $q_{a,b}^{SP}$'s
are not all equal, the solution of the saddle point equations is
degenerate, and one must average\footnote{Since the saddle point
solution is degenerate, one must sum over the solutions in both numerators
and denominators of~(\ref{replica10}).} over all degenerate solutions, or
alternatively all permutations of $a$ and $b$. The correct formula is
then

\begin{equation}
\overline{<\exp(\frac{y\beta^2}{N} \sum_i \sigma^{(1)}_i \sigma^{(2)}_i)>}
=\frac{2}{n(n-1)}\sum_{a<b} \exp(y \beta^2 q_{a,b}^{SP})\;,
\end{equation}
where $n(n-1)/2$ is the number of $q_{a,b}$ variables.  This implies
the following direct relation between the disorder averaged $P(q)$ and
the value of $q_{a,b}$ at the saddle point

\begin{eqnarray}
\label{replica3x}
P(q)&=&\overline{P(q)_{\myJ}}=\overline{<\delta(q-\frac{1}{N} \sum_i \sigma^{(1)}_i \sigma^{(2)}_i)>}\\
\nonumber
&=&\frac{2}{n(n-1)}\sum_{a<b} \delta(q-q_{a,b}^{SP})\;.
\end{eqnarray}

The distribution $P_{\myJ}(q)$, namely before disorder average, has
an interpretation in terms of pure states.  In the rigorous
formulation of statistical physics (without disorder), pure states
give a precise definition of thermodynamic phases directly in the
infinite volume limit (see~\cite{MaPaRiTeRULo} for a discussion from a
physicist point of view).  Pure states have the following properties:

\begin{itemize}

\item The clustering property: Inside a pure 
state, spin correlation functions factorize when the distance
goes to infinity.  For example 
if ${\alpha}$ is a pure state, one has
$<\sigma_x \sigma_y>_{\alpha}
\to<\sigma_x >_{\alpha}<\sigma_y>_{\alpha}$, as ${|x-y|\to\infty}$.

\item  Every translationally invariant state is a convex linear combination of pure states.
This means that for any $A$, one has
$<A>=\sum_{\alpha}w_{\alpha} <A>_{\alpha}$, where the weights
$w_{\alpha}\geq 0$, with $\sum_{\alpha}w_{\alpha}=1$, are independent
of $A$.

\end{itemize}

In a ferromagnet at temperature $T<T_c$, there are two $T$dependent
pure states corresponding to states with positive and negative
magnetizations respectively, that we can call the ``+'' and the ``-''
states. A general translationally invariant state is a linear
combination of these two pure states. For example, doing the infinite
volume limit (at zero magnetic field) by considering a set of finite
systems of increasing sizes with periodic boundary conditions, one has
$\omega_+=\omega_-=1/2$.

Let us assume (departing boldly from rigor) that the same
decomposition holds for finite systems in the SK model for every
disorder sample, namely that $<A>=\sum_{\alpha}w_{\alpha}
<A>_{\alpha}$, for any $A$, where the weights and the states are
(disorder) sample dependent.  We introduce $\myQ$, the overlap between
two pure states.

\begin{eqnarray}
\myQ_{\alpha,\beta}=
\frac{1}{N}\sum_i 
<\mysigmai>_{\alpha}
<\mytaui>_{\beta}\;.
\end{eqnarray}

In a ferromagnet $\myQ_{+,+}=\myQ_{-,-}=<M>^2$, and
$\myQ_{+,-}=\myQ_{-,+}=-<M>^2$.  We now proceed to show that the overlap
between pure states is related to the clone overlap introduced
in~(\ref{overlap}). We do this by considering successive moments of
$q$.
For the first moment, we have

\begin{eqnarray}
<q>&=&\frac{1}{N}\sum_i <\mysigmai\mytaui>\\
\nonumber
&=&\frac{1}{N}\sum_i <\mysigmai><\mytaui>\;,
\end{eqnarray}
since $\mysigma$ and $\mytau$ are independent systems. Introducing the pure states, this gives

\begin{eqnarray}
<q>   &=&\frac{1}{N}\sum_i \sum_{\alpha,\beta}
\omega_{\alpha}\omega_{\beta}
<\mysigmai>_{\alpha}
<\mytaui>_{\beta}\\
\nonumber
&=& \sum_{\alpha,\beta} \omega_{\alpha}\omega_{\beta}
\myQ_{\alpha,\beta}\;.
\end{eqnarray}

For the second moment we have

\begin{eqnarray}
<q^2>&=&\frac{1}{N^2}\sum_i\sum_j <\mysigmai\mysigmaj\mytaui\mytauj>
\\
\nonumber
&=&\frac{1}{N^2}\sum_{i,j} <\mysigmai\mysigmaj><\mytaui\mytauj>\\
\nonumber
&=&\frac{1}{N^2}\sum_{i,j} \sum_{\alpha,\beta}
\omega_{\alpha}
\omega_{\beta}
<\mysigmai\mysigmaj>_{\alpha}
<\mytaui\mytauj>_{\beta}\\
\nonumber
&\approx&\frac{1}{N^2}\sum_{i,j} \sum_{\alpha,\beta}
\omega_{\alpha}
\omega_{\beta}
<\mysigmai>_{\alpha}
<\mysigmaj>_{\alpha}
<\mytaui>_{\beta}
<\mytauj>_{\beta}\;,
\end{eqnarray}
where we assumed that, since the states $\alpha$ and $\beta$ are pure
states, one has $<\sigma_i \sigma_j>_{\alpha} \approx
<\sigma_i>_{\alpha} <\sigma_j>_{\alpha}$ (We pretend that all
points are far apart. This is clearly not correct
for $i=j$ but the error is negligible for large $N$).  Finally

\begin{eqnarray}
<q^2> &\approx& \displaystyle\sum_{\alpha,\beta} \omega_{\alpha}\omega_{\beta}
\myQ_{\alpha,\beta}^2\;.
\end{eqnarray}

In general one has, for any integer $r$,

\begin{eqnarray}
<q^r> &\approx& \displaystyle\sum_{\alpha,\beta} \omega_{\alpha}\omega_{\beta}
\myQ_{\alpha,\beta}^r\;,
\end{eqnarray}
namely (still with disorder dependent $\omega_{\alpha}$'s and $\myQ_{\alpha,\beta}$'s)

\begin{eqnarray}
<\delta(q-\frac{1}{N}\sum_i \mysigmai \mytaui)>   
   &= \displaystyle\sum_{\alpha,\beta} \omega_{\alpha}\omega_{\beta}
\delta(q-\myQ_{\alpha,\beta})\;.
\label{replica7}
\end{eqnarray}

This is a remarkable relation between a quantity relative to pure
states and a quantity that is directly accessible, e.g.~with Monte Carlo
simulation. This relation is quite useful since there is no simple way
to characterize pure states for spin glasses, in contrast to Ising
ferromagnets where a pure state can be selected by simply applying an
infinitesimal constant magnetic field of suitable sign.

\subsection{Solution of the model}
\label{subsec:10}

We look for the absolute minimum of $\myA(\{q_{a,b}\})$, namely the
lowest stable
solution of $\partial \myA/\partial q_{a,b}=0$ for all $q_{a,b}$ with
generic $n$.  The solution is then to be analytically continued to the
limit $n\to0$.   One proceeds
heuristically by first making an ansatz for the matrix $q_{a,b}$ and
generic $n$, involving a few parameters $x_i$ and $q_i$.  Very
surprising at first sight, the correct saddle point solution is a
maximum of $\myA$ with respect to the $x_i$'s and $q_i$'s (and not a
minimum), and if several maxima are found, one should usually take the
largest
.  Once a candidate solution is found, one should check that
this solution is stable (that the Hessian has only nonnegative
eigenvalues). If no satisfactory solution is found, one try a more
general ansatz (This is indeed a heuristic procedure).

\begin{figure}
\centering
\includegraphics[width=0.414\textwidth,angle=270]{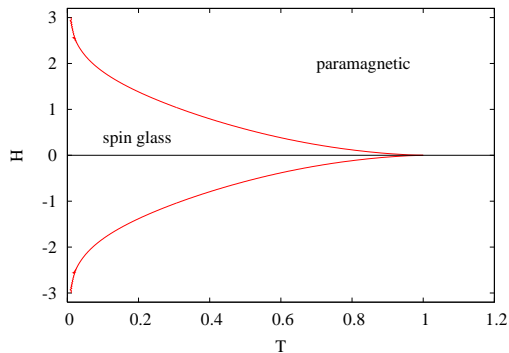}
\caption{Phase diagram of the SK model in the $T-H$ plane.
The paramagnetic RS solution is valid outside of the area delimited by the AT lines
and the $H$ axis. 
The complementary region is the spin glass phase, where the $\infty$-RSB ansatz
gives the correct result.}
\label{fig:2}       
\end{figure}

The simplest ansatz is the replica symmetric (RS) ansatz~\cite{ShKi},
namely all $q_{a,b}=q_0$, and accordingly $P(q)=\delta(q-q_0)$. This
ansatz has a (ferromagnetic) solution in the whole $T>0$ half-plane,
with $q_0=0$ and zero magnetization for $H=0$, and $q_0>0$ and nonzero
magnetization for $H\neq 0$. At zero magnetic field and $T<1$, this
ansatz has another solution with $q_0\neq 0$ that should in principle
be selected since it has a higher free energy than the first
solution. However detailed investigation~\cite{AlTh} of the
eigenvalues of the Hessian shows that neither solution is a maximum of
the integrand in~(\ref{replica11}), and accordingly both should be
rejected, when the absolute value of $H$ is below the so called AT
line that starts at some nonzero $H$ at $T=0$ and reaches the $H=0$
axis at $T_c=1$ (see Fig~\ref{fig:2}).  In the rest of the half $T>0$
plane, the paramagnetic RS solution is stable, which means
heuristically that it is the correct solution.

One is thus led to a more general, non replica symmetric, ansatz.  The
correct ($\infty$-RSB) ansatz was found by
Parisi~\cite{Pa1,Pa4}.  It is a hierarchical solution that goes
in an infinite number of steps.  The first step is the so called ``one
step RSB ansatz''.  The $n$ replica are partitioned in $n/x_1$ blocks
of $x_1$ replica.  At this level,  $x_1$ is an integer and
$x_1$ divides $n$.  One assumes that $q_{a,b}=q_1$ if $a$ and $b$ are
in the same block, and that $q_{a,b}=q_0$ if they do not.  With this
rule we have $-\frac{n}{2}(1-x_1)$ elements $q_{\alpha,\beta}$ 
(for $a>b$) with
value $q_1$, and $-\frac{n}{2}(x_1-n)$ elements with value $q_0$.  If
we plug this information into~(\ref{replica3x}), and do formally the
limit $n\to 0$, this gives

\begin{equation}
P(q)=x_1\delta(q-q_0)+(1-x_1)\delta(q-q_1)\;,
\end{equation}
which means that $P(q)$ has two peaks of weights $x_1$ and $1-x_1$
respectively. Clearly this result only makes sense if $x_1\in
[0,1]$. We started with a $n \times n$ matrix subdivided in $x_1
\times x_1$ blocks, and then let $n \to 0$.  We have to admit that in
this strange limit, the integer $x_1$ became a real variable $\in [0,1]$.

In the next step, the so called ``two steps RSB ansatz'', each diagonal block of
size $x_1$ is divided in $x_1/x_2$ sub blocks of size $x_2$ and one assumes
that now $q_{\alpha,\beta}=q_2$ in the diagonal sub-blocks, and is
unchanged elsewhere.

Performing   this procedure $k$ times, we introduce a subdivision $
n=x_0>x_1>x_2>\cdots>x_{k+1}=1$, with integer $x_0,x_1,\ldots,x_k$,
$x_0/x_1$, $x_1/x_2,\ldots,x_{k-1}/x_k$ and to a set of values for the
overlaps $q_0,q_1, \ldots,q_k$. Then

\begin{equation}
P(q)=\sum_{i=0}^k (x_{i+1}-x_{i})\delta(q-q_i)\;.
\label{replica5}
\end{equation}

In order for this formula to makes sense, we have to assume that in
the $n\to 0$ limit, the $x_i$ variables became $
0=x_0<x_1<x_2<\cdots<x_{k+1}=1$.  Assuming now that the $q_i$'s form
an increasing sequence, one has

\begin{equation}
\int_0^{q_m}P(q)dq =\sum_{i=0}^{m} (x_{i+1}-x_i)=x_{m+1} \qquad m=0,1,\ldots, k\;.
\end{equation}

In the $k\to\infty$ limit, assuming that the increasing sequence
$q_i$'s converges to a continuous function $q(x)$, with an inverse
x(q), this equation becomes

\begin{equation}
\int_0^{q}P(q')dq' =x(q), \qquad P(q)=\frac{dx}{dq}\;.
\end{equation}

We have similarly the equation, for any integer $r$,

\begin{equation}
-\frac{2}{n} \sum_{a\neq b} q_{a,b}^r =
\sum_{i=0}^k (x_{i+1}-x_{i})q_i^r \sim \int_0^1 dx q(x)^r\;.
\end{equation}

Finally
$\myA(q(x))$ is expressed as a (complicate) functional of $q(x)$,
that should be maximized with respect to variations of the function
$q(x)$ in order to give the free energy $f(T)$:

\begin{equation}
 -f(T) = \frac{\beta}{4}\,\Bigl( 1 - 2\,q(1) + \int_0^1dx\, q^2(x)
 \Bigr) + \int_{-\infty}^{+\infty} \frac{d y}{\sqrt{2 \pi q(0)}}
\exp\left(-\frac{(y-H)^2}{2\,q(0)}\right)\phi(0,y)\;,
\label{zeequation}
\end{equation}
where $\phi(0,y)$ is the solution, evaluated at $x=0$, 
of the equation
\begin{equation}
\partial_x\phi(x,y)=-\frac{\partial_x{q}(x)}{2}\,
\Bigl[
\partial^2_y\phi(x,y)+\beta\,x\,(\partial_y\phi(x,y))^2
\Bigr]\;,
\end{equation}
with the boundary condition
\begin{equation}
\phi(1,y)= \beta^{-1}\log\left(2\cosh \beta y\right)\;.
\label{Phi1}
\end{equation}

One can show~\cite{DeDoKo} that the ($\infty$-RSB) solution is stable
in the whole region where the paramagnetic $RS$ solution is unstable
.  This means (heuristically) that we have found the correct solution
for all $H$ and $T>0$.  There is however no known close form solution
neither for $q(x)$~\footnote{From now on, $q(x)$ denotes the solution
of the saddle point equations $q^{SP}(x)$.} nor for the corresponding
free energy $f(T)$, and most articles in the literature use
approximate estimates of $q(x)$, usually based on the truncated
Hamiltonian (see next section), valid for $T\approx T_c$. Precise
estimates can be obtained however either by an expansion of the
functional in powers of $T_c-T$, or by numerical methods
(see~\cite{CrRiz} for recent very precise estimates in the $H=0$ case,
and~\cite{CrRizTe} in the $H\neq0$ case).

\begin{figure}
\begin{picture}(100,100)(-80,-30)  
\put(0,0){\vector(1,0){100}}  
\put(0,0){\vector(0,1){60}}
\put(60,-3){\line(0,1){6}}
\put(-10,60){q}
\put(90,-10){x}
\put(60,-10){1}
\put(10,-10){$x_{min}$}
\put(-30,40){$q_{EA}$}
\put(20,40){\line(1,0){40}}
\put(10,20){\line(1,2){10}}
\put(-30,20){$q_{min}$}
\put(0,20){\line(1,0){10}}
\put(140,0){\vector(1,0){70}}  
\put(140,0){\vector(0,1){60}}
\put(110,40){P(q)}
\put(210,-10){q}
\put(160,20){\line(1,0){20}}  
\put(180,0){\line(0,1){50}}   
\put(160,0){\line(0,1){30}}   
\put(180,-10){$q_{EA}$}
\put(150,-10){$q_{min}$}
\end{picture}
\caption{Schematic representation of the $\infty$-RSB solution 
for $H\neq0$: $q(x)$ as a function of $x$ and $P(q)$ as a function of
$q$,  with $q_{EA}=q_{max}$. In the actual solution the increasing
portion of $q(x)$ and the corresponding continuum in $P(q)$ are not
straight.}
\label{figure20}
\end{figure}
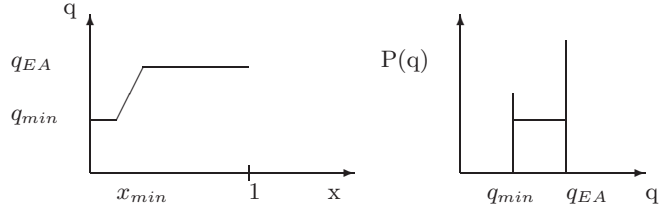

For $H\neq 0$, the continuous non decreasing function $q(x)$ behaves
as follows (see Fig~\ref{figure20}): There exist values $0\leq
x_{min}\leq x_{max}\leq 1$ such that $q(x)=q_{min}\geq 0$ for $x\in
[0,x_{min}]$, $q(x)$ increases for $x\in [x_{min},x_{max}]$, and
$q(x)=q_{max}$ for $x\in [x_{max},1]$.  Accordingly the function
$P(q)$ has two delta function peaks located at $q(x)=q_{min}$ and
$q(x)=q_{max}$ respectively, and is nonzero between.  The transition
on the AT line is continuous, with $q_{min}\to q_{max}$, as the AT
lined is approached.

The value $q_{max}$ is interpreted as $q_{EA}$, 
the overlap between two configurations in the same pure
state

\begin{equation}
q_{EA}=\frac{1}{N}\sum_i <\sigma_i>_{\alpha}<\sigma_i>_{\alpha}\;.
\end{equation}

One can show that $q_{EA}$ is independent of $\alpha$. This
definition is in agreement with the one given in the
introduction~(\ref{qEA}) thanks to the wide separation of time scales
in the model: A given configuration stays a very long time in the same
pure state, and accordingly the correlation function $C(t)$ has a
plateau of height $q_{EA}$.

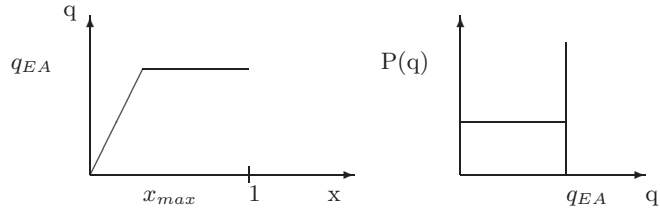
\begin{figure}
\begin{picture}(100,100)(-80,-30)  
\put(0,0){\vector(1,0){100}}  
\put(0,0){\vector(0,1){60}}
\put(60,-3){\line(0,1){6}}
\put(-10,60){q}
\put(90,-10){x}
\put(60,-10){1}
\put(20,-10){$x_{max}$}
\put(-30,40){$q_{EA}$}
\put(20,40){\line(1,0){40}}
\put(0,0){\line(1,2){20}}
\put(140,0){\vector(1,0){70}}  
\put(140,0){\vector(0,1){60}}
\put(110,40){P(q)}
\put(210,-10){q}
\put(140,20){\line(1,0){40}}  
\put(180,0){\line(0,1){50}}   
\put(180,-10){$q_{EA}$}
\end{picture}
\caption{Schematic representation of the 
$\infty$-RSB solution for $H=0$: $q(x)$ as a function of $x$ and
$P(q)$ as a function of $q$,  with $q_{EA}=q_{max}$.  In the
actual solution the increasing portion of $q(x)$ and the corresponding
continuum in $P(q)$ are not straight. In this $H=0$ case, $P(q)$ is
symmetric in $q$.  Only the $q>0$ part of $P(q)$ is represented here.}
\label{figure21}
\end{figure}

As $H\to 0$, $q_{min}\to 0$ and the corresponding peak disappears.
When $H$ is exactly zero, $P(q)$ which is nonzero for $0<q_{min}\leq
q \leq q_{max}$ for $H\neq 0$ becomes symmetric (obviously the two
limits $N\to\infty$ and $H\to 0$ do not commute). At $H=0$, P(q) is
made of two delta peaks located at $\pm q_{max}=\pm q_{EA}$ with a
continuum between. As shown in~\cite{CrRiz}, and in contradiction with
many drawings in the literature (and Fig~\ref{figure21}), $P(q)$ has a
minimum at a nonzero value of $q$ (at least as soon as
$T<0.96\ldots$) and behaves like $a+bq^2+c|q|^3+\cdots$, with $c\neq
0$ for small $q$.  When $T\to T_c$, $q_{EA}\to 0$, namely the
transition is continuous.

\subsection{Properties of the solution}
\label{subsec:11}

Among the fascinating features of Parisi solution, is ultrametricity~\cite{MePaVi84}.
Taking three clones, one can show
that

\begin{eqnarray}
P(q_{1,2},q_{2,3},q_{3,1})&=&
<\delta(q_{1,2}-\sum_{i=1}^N \sigma_i^{(1)}\sigma_i^{(2)})
 \delta(q_{2,3}-\sum_{i=1}^N \sigma_i^{(2)}\sigma_i^{(3)})\\
\nonumber
& & \delta(q_{3,1}-\sum_{i=1}^N \sigma_i^{(3)}\sigma_i^{(1)})>\\
\nonumber
&=&\frac{1}{n(n-1)(n-2)}\sum_{a,b,c}^{a\neq b, b\neq c, c\neq a} \delta(q_{1,2}-q_{a,b})\\
\nonumber
&& \delta(q_{2,3}-q_{b,c})\delta(q_{3,1}-q_{c,a})\;.
\end{eqnarray}

The $\infty$-RSB solution for the $q_{a,b}$'s has the property that
this probability is zero unless the three overlaps
$q_{1,2},q_{2,3},q_{3,1}$ satisfy the ultrametricity constraints: at
least two overlaps are equal, and the third is smaller or equal to
their common value. Ultrametricity generalizes to pure states, where
one defines a distance between pure states $\alpha$ and $\beta$, as

\begin{eqnarray}
d_{\alpha,\beta}&=&\frac{1}{N}\sum_i (<\sigma_i>_\alpha-
<\sigma_i>_\beta)^2,\\
\nonumber
&=&\myQ_{\alpha,\alpha}+\myQ_{\beta,\beta}-2\myQ_{\alpha,\beta}
= 2 (q_{EA}-\myQ_{\alpha,\beta})\;,
\end{eqnarray}
since all self overlaps are equal to $q_{EA}$. 

Another feature~\cite{MePaVi84} of the Parisi solution is that some intensive
observables are not self averaging~\footnote{For a discussion of
non-self averaging in other random systems, see~\cite{WiDo}.} namely
some thermodynamic averages $<A>$ retain a (disorder) sample
dependence even in the large $N$ limit, i.e.  $\lim_{N\to\infty}
(\overline{<A>^2}-{\overline{<A>}}^2)\neq 0$.  Among self averaging
quantities are the internal energy, the free energy, the magnetization
and the Edwards--Anderson order parameter $q_{EA}$.  The order
parameter distribution function $P_J(q)$ is an example of a non self
averaging, quantity. The general rule is that observables that do not
involve correlations between different pure states are self averaging,
those that involve such correlations are not self averaging.

In order to show that the order parameter distribution function
$P_J(q)$ is not self averaging, one considers four clones, and the
probability $P_J(q_{1,2},q_{3,4})$. Clearly
$P_J(q_{1,2},q_{3,4})=P_J(q_{1,2}) P_J(q_{3,4})$, since clones are non
interacting.  However, we will show that

\begin{equation}
\overline{P_J(q_{1,2}) P_J(q_{3,4})}
\neq
\overline{P_J(q_{1,2})}\
\overline{P_J(q_{3,4})}\;,
\label{guerra}
\end{equation}
which implies that $P_J(q)$ is not self averaging, since a self
averaging $P_J(q)$ would not depend on $J$ for a large system.  The
evaluation of $\overline{P_J(q_{1,2},q_{3,4})}$ is done by considering
moments of the distribution

\begin{eqnarray}
&&\int dq_{1,2}\ q^r_{1,2}
\int dq_{3,4}\ q^s_{3,4}
\overline{P_J(q_{1,2}) P_J(q_{3,4})}
=\overline{<q^r_{1,2} q^s_{3,4}>}\\
\nonumber
&=&\frac{1}{n(n-1)(n-2)(n-3)}
\sum_{\substack{a,b,c,d\\{\rm all\ different}}} q^r_{a,b} q^s_{c,d}\;,
\end{eqnarray}
with integer $r$ and $s$.

We now use the very powerful property of replica equivalence, namely
the observation that in Parisi solution one has:

\begin{equation}
\sum_{b} q_{a,b} \qquad {\rm independent \ of}\ a\;.
\end{equation}

In order to use this property, one notes that
.

\begin{eqnarray}
\sum_{\substack{a,b,c,d\\{\rm all\ different}}} q^r_{a,b} q^s_{c,d}
&=&\sum_{a,b}\sum_{c,d} q^r_{a,b} q^s_{c,d}
-4\sum_{a,b,d} q^r_{a,b} q^s_{a,d}\\
\nonumber
&+&2\sum_{a,b} q^r_{a,b} q^s_{a,b}\;.
\end{eqnarray}

Since the sum $\sum_{a,b,d} q^r_{a,b} q^s_{a,d}$ is unrestricted, one can write it as

\begin{eqnarray}
\sum_{a,b} q^r_{a,b} \sum_d q^s_{a,d}&=&\sum_{a,b} q^r_{a,b} \frac{1}{n}\sum_{c,d} q^s_{c,d}\;.
\end{eqnarray}

In the $n\to 0$ limit, one obtains

\begin{eqnarray}
\overline{<q^r_{1,2} q^s_{3,4}>}
=\frac{1}{3} \overline{<q^{r+s}_{1,2}>}
+\frac{2}{3} \overline{<q^r_{1,2}>}\ \overline{<q^s_{3,4}>}\;,
\label{Guerra1}
\end{eqnarray}

\begin{eqnarray}
\overline{ P_J(q_{1,2}) P_J(q_{3,4})}
=\frac{1}{3} \overline{P_J(q_{1,2})}\delta(q_{1,2}-q_{3,4})
+\frac{2}{3} \overline{P_J(q_{1,2})}\ \overline{P_J(q_{3,4})}\;,
\label{Guerra2}
\end{eqnarray}
in agreement with~(\ref{guerra}).
It is interesting that the above relation can be proven rigorously 
~\cite{Gu1} (the proof given in this paper is for the $r=s=2$ case).

The method used by Parisi to solve the SK model is far from
mathematical rigor, to say the less, and it took time to convince the
skeptics that it does give the correct result.  Very recently however
beautifully rigorous results have been obtained. One is a rigorous
proof~\cite{GuTo} that the free energy density of the SK model has a
well defined limit as $N\to\infty$, a long awaited result that is far
from trivial for a disordered model with infinite connectivity. Then
the method was extended to show that the free energy of the Parisi
solution is a lower bound on the free energy of the model~\cite{Gu2},
and finally to prove that it is equal to the free energy of the
model~\cite{Tala}. More recently a proof appeared that the AT line is
indeed the boundary of the paramagnetic replica symmetric
region~\cite{Gu3}.

\subsection{Critical exponents}
\label{subsec:12}

At zero magnetic filed, the transition at $T_c$ is of second order and
one can define critical exponents. The order parameter is the mean
overlap $\overline{<q>}$, and one has

\begin{eqnarray}
\overline{<q>}&\propto& (T_c-T)^{\beta} \qquad  T<T_c 
\qquad {\rm with} \qquad \beta = 1\;, \\
\nonumber
&=&0 \hskip 2cm  T>T_c \;,\\ 
\nonumber
\chi_{SG}&=& N  \overline{<q^2>}\;,\\
\nonumber
&\propto& (T_c-T)^{-\gamma} \qquad  T>T_c 
\qquad {\rm with} \qquad \gamma = 1\;,\\
\nonumber
&= & \infty \hskip 1.8 cm  T<T_c\;. 
\end{eqnarray}

The values of the critical exponents are $\alpha=-1$ (the specific
heat has a cusp at the critical temperature), $\beta=1$, $\eta=0$ (the
mean field value), $\gamma=1$, and $\nu=1/2$. Hyper scaling holds
provided one uses the value $d=6$ of the space dimension in the
formula. This is related to the fact that $d=6$ is the upper critical
dimension of the replica field theory.

\subsection{The truncated model}
\label{subsec:13}

We have seen that for $H=0$ the high $T$ solution has $q_{a,b}=0\
\forall a,b$. It
is accordingly natural to expand $\myA$ in powers of $q_{a,b}$
in~(\ref{replica11}). Only the terms up to the order four are usually
kept, as one can show that this gives the correct critical behavior
and the correct qualitative description of the low $T$ phase.  This truncated
model has been very useful historically. It is also quite useful in
order to study finite size effects (See Subsec.~\ref{subsec:20}).

The trace over the spins is straightforward.  For a given replica
index, the trace is equal to zero for an odd power of the replicated
spins and to two for an even power.  Neglecting powers of $q_{a,b}$ higher
than four, one obtains the truncated model (written for $H=0$):

\begin{equation}
\beta f_N=-\ln{2}-\frac{\beta^2}{4}
-\lim_{n\to0}\frac{1}{n N}\ln\int \Biggl[\prod_{a< b}\sqrt{\frac{N}{2\pi\beta^2}}\
d\qtilda_{ab}
\Biggr]\exp(-N \beta\myL[\qtilda ])\;.
\label{trunc1}
\end{equation}

\begin{eqnarray}
\label{trunc2}
\beta\myL[\qtilda]&=&\frac{\tau}{2}\sum_{a, b} \qtilda_{a,b}^2
-\frac{1}{6}\sum_{a,b,c} \qtilda_{a,b}\ \qtilda_{b,c}\ \qtilda_{c,a}\\
\nonumber
&-&\frac{1}{8}\sum_{a,b,c,d} \qtilda_{a,b}\ \qtilda_{b,c}\ \qtilda_{c,d}\ \qtilda_{d,a}
+\frac{1}{4}\sum_{a,b,c} \qtilda_{a,b}^2\ \qtilda_{a,c}^2
-\frac{1}{12}\sum_{a,b} \qtilda_{a,b}^4
+\cdots\;,
\end{eqnarray}
where $\tau=(T^2-1)/2$.
We use the notation $\qtilda=q\beta^2= q/T^2$ in order to
simplify the formula, and write unrestricted sums over the replica
indexes, considering $\{q_{a,b}\}$ as a symmetric matrix with zero
elements on the diagonal.  For $T>T_c$, $\tau>0$ and the paramagnetic solution is stable
as it should.

A further simplification, that leads to simpler analytical
computations, has been proposed by Parisi in one of his seminal
articles~\cite{Pa3}.  It amounts to replace~(\ref{trunc2}) by

\begin{eqnarray}
\beta \myL[\qtilda]&=&\frac{\tau}{2}\sum_{a,b} \qtilda_{a,b}^2
-\frac{1}{6}\sum_{a,b,c} \qtilda_{a,b}\ \qtilda_{b,c}\ \qtilda_{c,a}
-\frac{y}{8}\sum_{a,b} \qtilda_{a,b}^4\;,
\label{trunc3}
\end{eqnarray}
keeping the only fourth order term that is responsible for the replica
symmetry breaking and using the value $y=2/3$. This is the Parisi
approximation of the truncated model, sometimes called the truncated
model, sometimes the
reduced model. In most cases this approximation has only mild
effects, changing the numerical value of some coefficients. In a few
cases however, it gives qualitatively wrong results.

\subsection{Some variant of the model}

Two variants of the Sherrington--Kirkpatrick model are worth
mentioning.  The first is the spherical Sherrington--Kirkpatrick
model\index{Spherical Sherrington--Kirkpatrick model}, defined by the
same Hamiltonian~(\ref{Hamiltonian}) as the original model, but
with continuous spins $\sigma_i$ obeying the spherical constraint

\begin{equation}
\sum_i \sigma^2_i = N\;.
\end{equation}

It can be exactly solved~\cite{spherical} in the thermodynamic limit
without the use of replica.  It can also be solved using the replica
trick, and both methods give the same results. The physics of the
spherical model is however quite different from the one of the
original model: It is paramagnetic for all $T\geq 0$ and $H\neq 0$ but
for the region $H=0$ and $T\leq 1$, where the RS ansatz with nonzero
$q_0$ gives the correct solution. On the other hand, much has been
learned of the dynamics of spin glass models from analytical
studies of the spherical SK model (see e.g.~\cite{Cu}).

The second variant is the p-spin model~\cite{EG}, where all combinations of
$p$ spins are coupled together, with the Hamiltonian\index{p-spin model}

\begin{eqnarray}
H= - \sum_{1\leq i_1 < i_2 < \cdots < i_p \leq N} J_{i_1,\ldots,i_p} \ \sigma_{i_1} \cdots \sigma_{i_p} 
- H \sum_{i=1}^N \sigma_i\;,
\end{eqnarray}
where the $J_{i_1,\ldots,i_p}$'s are quenched independent identically
distributed random variables with zero mean and square deviation
$\overline{ J^2}=\displaystyle\frac{p!}{2 N^{p-1}}$.  The spins are
Ising spins $\sigma_i=\pm 1$.  The original SK model is recovered for $p=2$.
The $p\geq 3$ p-spin model has a very different physics than the SK
model: in the $H=0$ case, by decreasing the temperature from $T=\infty$
one encounters three successive phase transitions, a purely dynamical
transition, then a transition to a state with one-step RSB, and finally
a transition to a state with $\infty$-RSB.  For details seen
\cite{EG,MaRi}. One can  define a spherical p-spin model with
continuous spin variables obeying the spherical
constraint~\cite{CrSo,CrHoSo}. Many analytical results have
been obtained for the later model.  Some subtle qualitative
differences are found however between the original and spherical
p-spin models~\cite{MaRi}.

\section{Simulations techniques}

Spin glass simulations are very difficult, since on the one hand the
dynamics is very slow, and on the other hand one needs to perform the
simulation for many disorder samples. Both effects are stronger and
stronger as $T$ decreases and/or $N$ increases. The SK model is no
exception. It is furthermore much harder to simulate than finite
dimension spin glass models since one needs $O(N)$ operations to
update one spin variable, and not $O(1)$. The p-spins model is even harder
with $O(N^{p-1})$ operations to update a single spin. For a simulation
of the $p=3$ p-spin model, see~\cite{BGM}.

The best existing algorithm is called under various names such as
``Replica Monte Carlo'', ``Exchange Monte Carlo'' and ``Parallel
Tempering''~\cite{Partemp}\index{Parallel tempering algorithm}. This
algorithm is well known and consists in simulating $n_{PT}$ clones of
the system in parallel at temperatures $T_1 < T_2 < \cdots <
T_{n_{PT}}$ respectively. Two kind of Monte Carlo moves are performed
\begin{itemize}
\item Step 1 (the parallel Metropolis step): Metropolis 
update of each clone independently.
\item Step 2 (the exchange step): Conditionally exchange of 
the spin configurations of all pairs of clones, with a suitable
acceptance probability.
\end{itemize}

In both steps the acceptance probability is the usual Metropolis
acceptance probability.  If $\pi_x$ is the Boltzmann weight of state
$x$, with energy $E_x$, namely $\pi_x=\exp(-E_x/T)/\sum_y \exp(-E_y/T)$,
and $p_{x,y}$ the probability to go from state $x$ at time $t$ to
state $y$ at time $t+1$, one desires to fulfill the condition

\begin{equation}
\sum_x \pi_x p_{x,y} = \pi_y\;.
\end{equation}

This condition is called balance (stationarity in the mathematical
literature). Together with ergodicity (irreducibility in the
mathematical literature) and the more technical condition of
aperiodicity, this ensures that the Markov chains converges towards
the Boltzmann distribution~\cite{So,Zi,Yo}.

Balance is obtained by requiring that the probability to propose the move
$p^{(0)}_{x,y}$ and the probability to accept it $a_{x,y}$
satisfy~\footnote{I consider the general case $p^{(0)}_{x,y} \neq
p^{(0)}_{y,x}$, even if the equality usually holds.} (One has clearly
$p_{x,y} = p^{(0)}_{x,y} a_{x,y}$)

\begin{equation}
\frac{p^{(0)}_{x,y} a_{x,y}} {p^{(0)}_{y,x} a_{y,x}}=\min(1,\exp(-1/T(E_x-E_y))) \;.
\end{equation}

A common misconception is that after the exchange step the system is not at
equilibrium, and that this introduces a bias, that one may minimize by
making many parallel Metropolis steps (step 1) between each exchange
step. This is not correct, balance is enough to ensure convergence
towards the Boltzmann weight.

There is a considerable freedom in implementing this algorithm.  The
parallel Metropolis step is usually done by updating all spins, either
systematically (all spins are updated one after the other), or
randomly (choose one spin at random, then update it, then choose
another, \ldots). The exchange step is usually restricted to the
$n_{PT}-1$ pairs of clones with neighboring temperatures (the acceptance
rate of this exchange is higher), and can be done systematically
(update the pair at temperatures $T_1$ and $T_2$, then the pair at
temperatures $T_2$ and $T_3$, \ldots) or randomly. Step two clearly
takes no time as compared to step one. One usually alternates the two
steps, one parallel tempering step (PT step) consists accordingly of one parallel
Metropolis step followed by one exchange step.

The choice of the sequence $T_1 < T_2 < \cdots < T_{n_{PT}}$ leads
also to considerable freedom.  One constraint is that the acceptance
rate of the chain exchange (chain of energy $E$ at temperature $T$
with chain of energy $E'$ at temperature $T'$)

\begin{eqnarray}
\frac{a_{T,T'}}{a_{T',T}}&=&\min(1,\exp(-(E/T'+E'/T))/\exp(-(E/T+E'/T')))\\
\nonumber
&=&\min(1-\exp(-(E-E')(1/T-1/T')))\;,
\end{eqnarray}
should not be too small.  The acceptance rate is thus governed by the
combination $(T'-T)^2 dE/dT$, with $dE_N/dT=N de_N(T)/dT=Nc_N(T)$,
where $c_N(T)$ is the specific heat per spin, which is regular for all
temperatures (and weakly $N$ dependent) in the SK model. One
may~\cite{KeRe} adjust for every system size $N$ the number $n_{PT}$
and the values of the temperature of the clones in such a way that
$(T_{i+1}-T_i)^2 N c_N(T)$, $i=1,2,\ldots, n_{PT}-1$ is roughly
independent of $i$ and $N$. There is however no guaranty that this
choice is optimal, for example that it minimizes the statistical
errors.  A more satisfactory prescription is to maximize the number of
tunnelings, namely the number of times a given clone goes from the
minimal temperature to the maximal temperature (or the other way), as
proposed recently in~\cite{TrHuTr}.
.

\subsection{An example of Monte Carlo simulation}

In what follows I will  use data generated in collaboration
with Enzo Marinari~\cite{BiMa} for $H=0$ (with $N=64$ to $4096$,
$T\in[0.4, 1.325]$) and Barbara Coluzzi~\cite{BiCo} for $H\neq 0$ (for
$N$ up to $3200$).  For the purpose of these notes, I performed additional
small simulations in order to measure the number of tunnelings as a
function of $N$. I considered $N=64, 256$, and $1024$. The
temperatures of the clones are equidistant with $T=0.4, 0.425, 0.45,
\ldots, 1.325$ (here $n_{PT}=38$), with 128 disorder  samples (16 for $N=1024$),
performing $400000$  PT steps, starting from very well
equilibrated configurations.

\begin{figure}
\centering
\includegraphics[width=0.414\textwidth,angle=270]{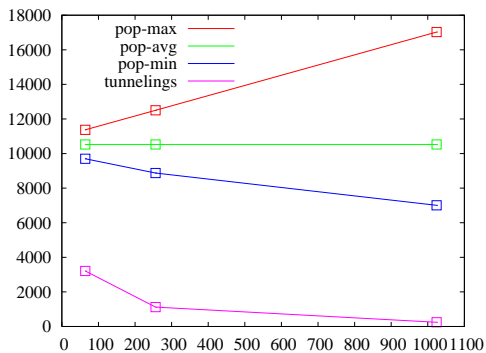}
\caption{From top to bottom: maximal (pop-max), average (pop-avg) and minimal
(pop-min) times spent by a clone at a given temperature as a function of
the number of sites.  Lines have been drawn between the points to
guide the eyes.  The line in the bottom is the average number of
tunnelings.}
\label{fig:1}       
\end{figure}

As shown in Fig.~\ref{fig:1}, the number of tunnelings is dramatically
decreasing as $N$ grows.  Figure~\ref{fig:1} gives another indicator of
the algorithm behavior: the spread of the distribution of times spent
by a given chain at each temperature.  With infinite statistics, each
chain should spend the same amount of time at each temperature. I have
measured for each disorder sample the maximal (pop-max) and minimal
(pop-min) time spent (in unit of PT steps) by each chain at each
temperature.  The average time (pop-avg) is equal to
$400000/n_{PT}=10526$.  These numbers are plotted in Fig.~\ref{fig:1} as a
function of $N$. The situation is clearly degrading as $N$ grows.
Analyzing the $N=4096$ data of~\cite{BiMa} (The results of a massive
numerical effort, with $520000$ PT steps and $\Delta T=0.0125$) one
finds pop-min as low as $16.5$.

\subsection{Comparison with theoretical expectations}

\begin{figure}
\centering
\includegraphics*[height=4.5cm,angle=270]{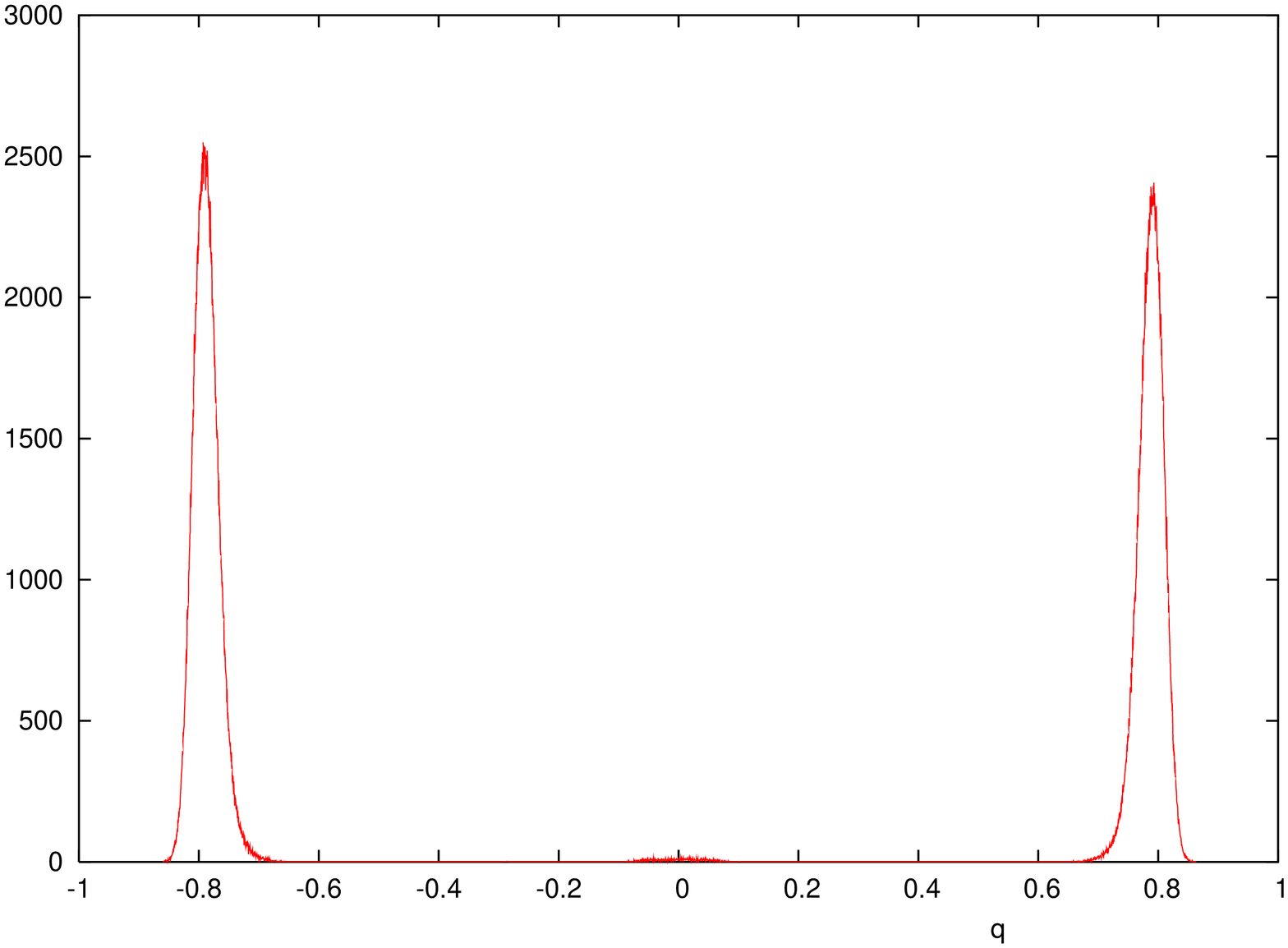}
\hskip 2cm
\includegraphics*[height=4.5cm,angle=270]{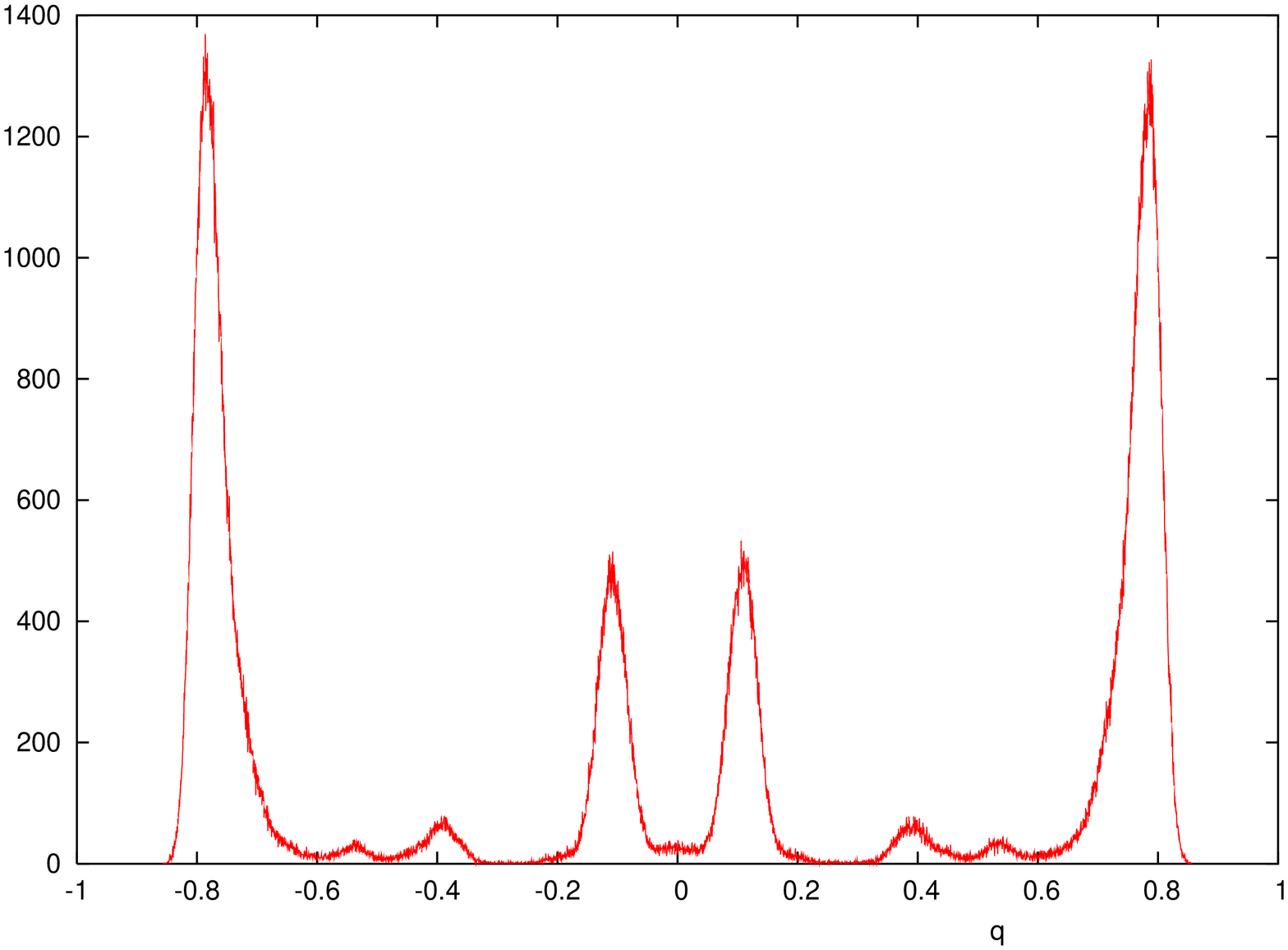}
\hskip -2cm
\vskip  1cm
\includegraphics*[height=4.5cm,angle=270]{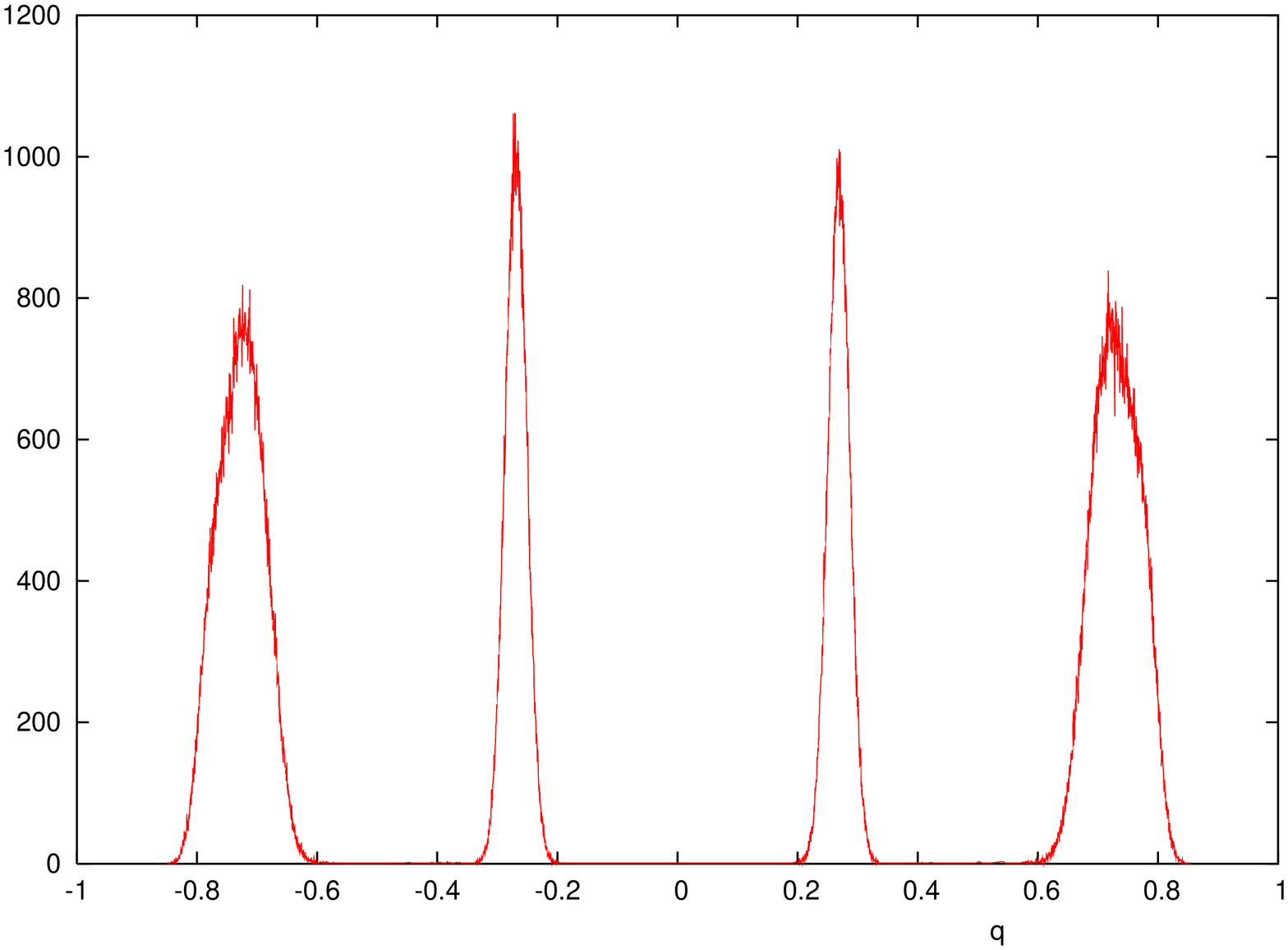}
\hskip 2cm
\includegraphics*[height=4.5cm,angle=270]{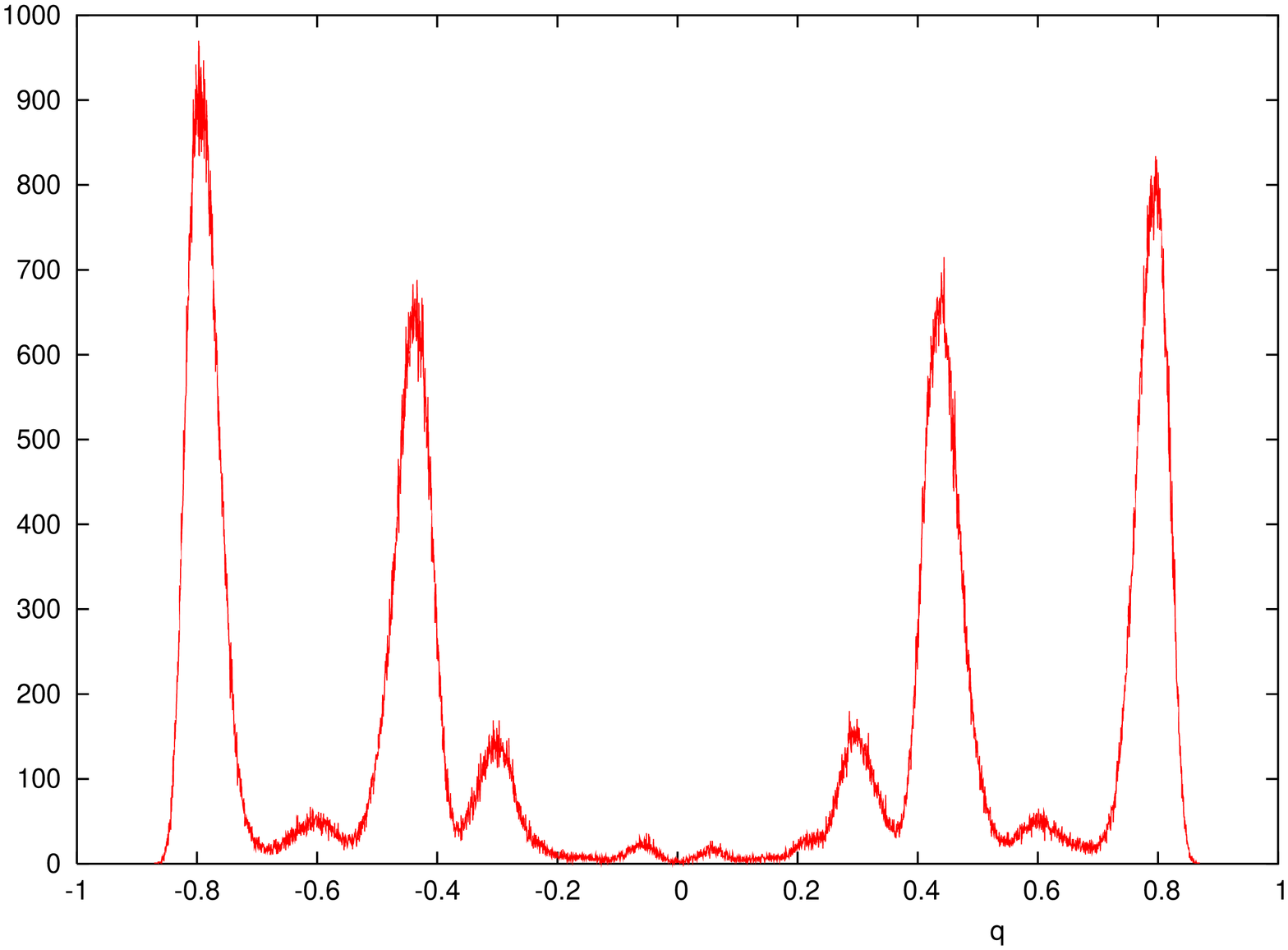}
\hskip -2cm
\vskip  1cm
\includegraphics*[height=4.5cm,angle=270]{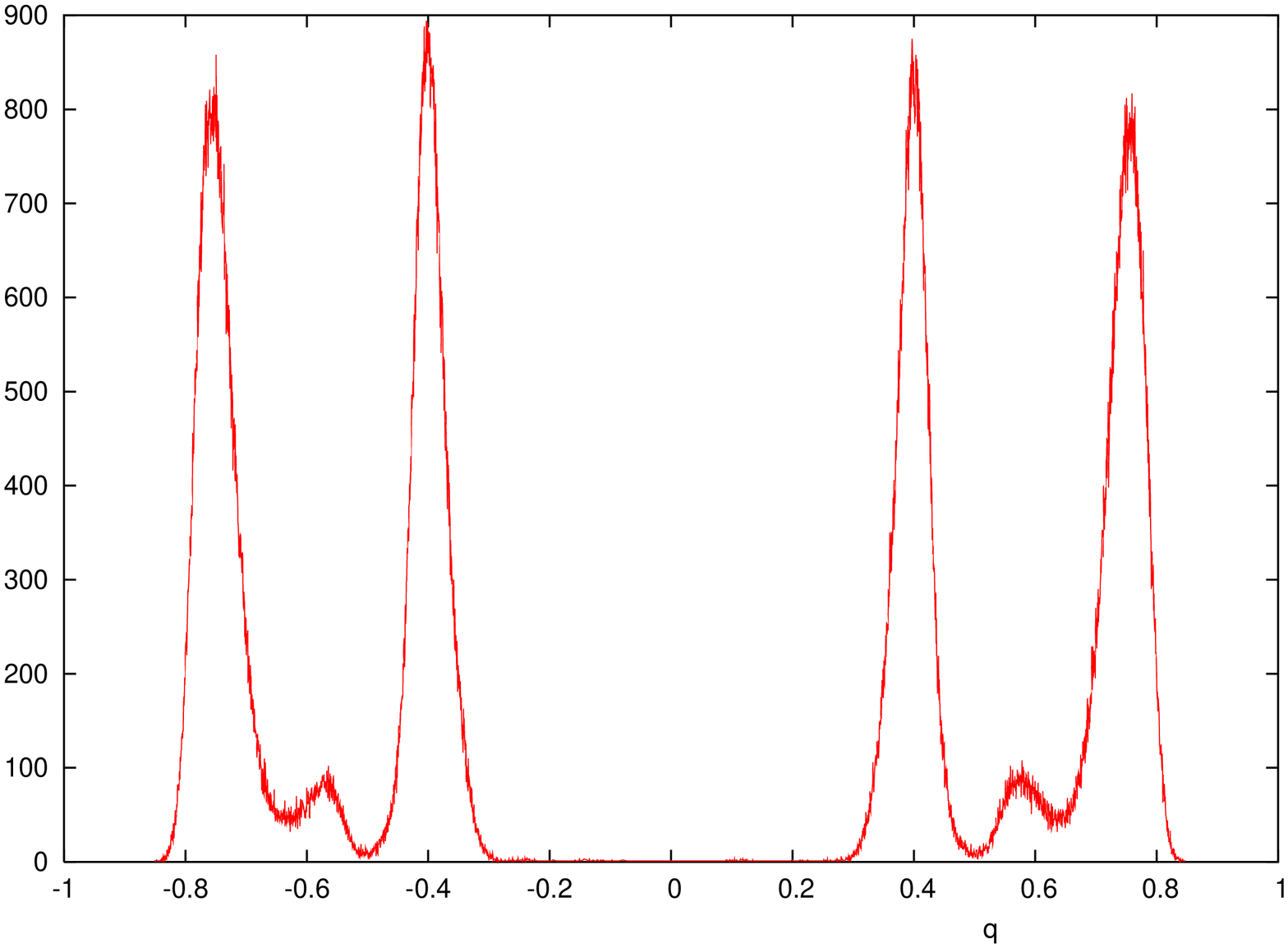}
\hskip 2cm
\includegraphics*[height=4.5cm,angle=270]{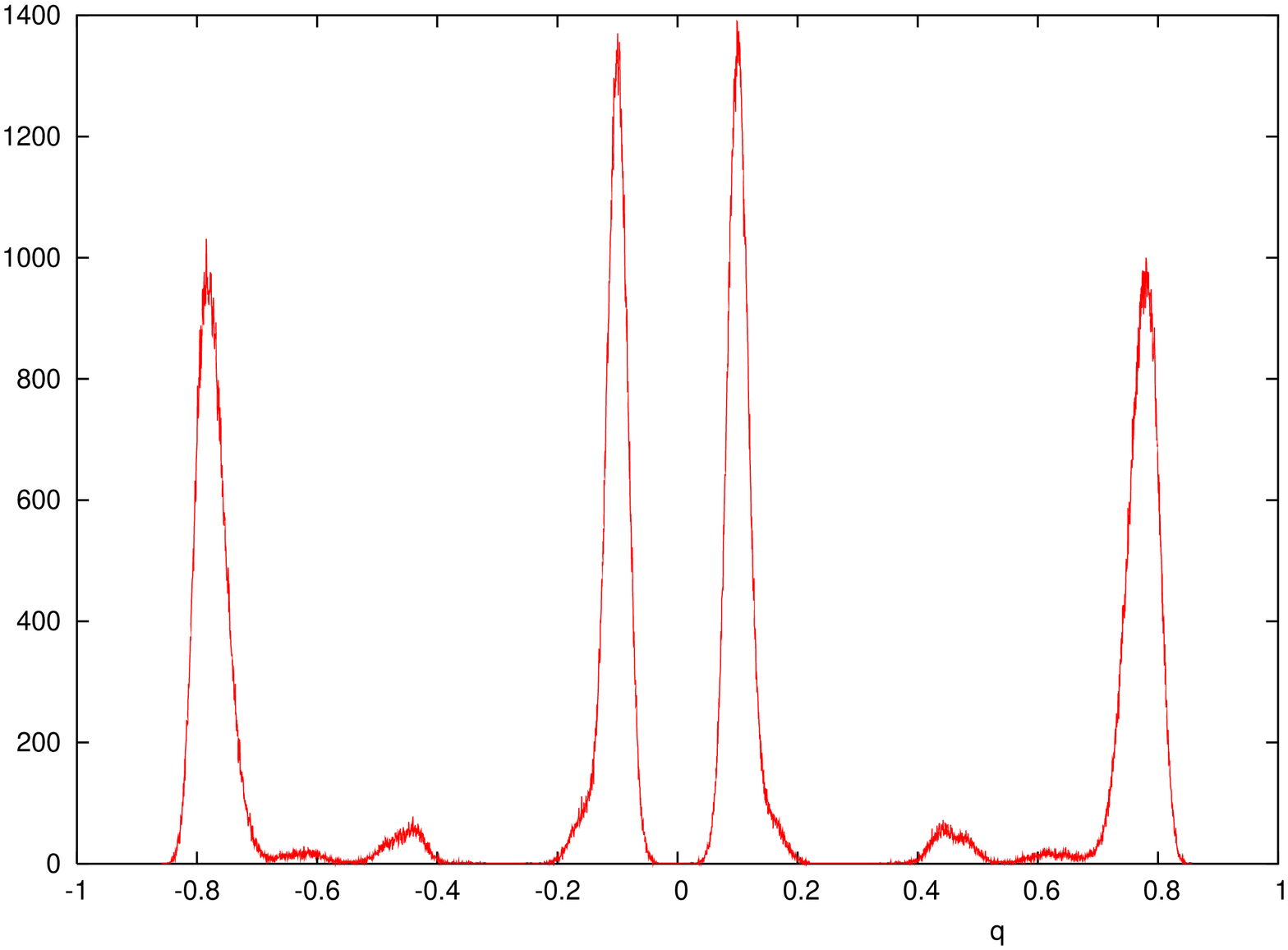}
\hskip -2cm
\vskip  1cm
\includegraphics*[height=4.5cm,angle=270]{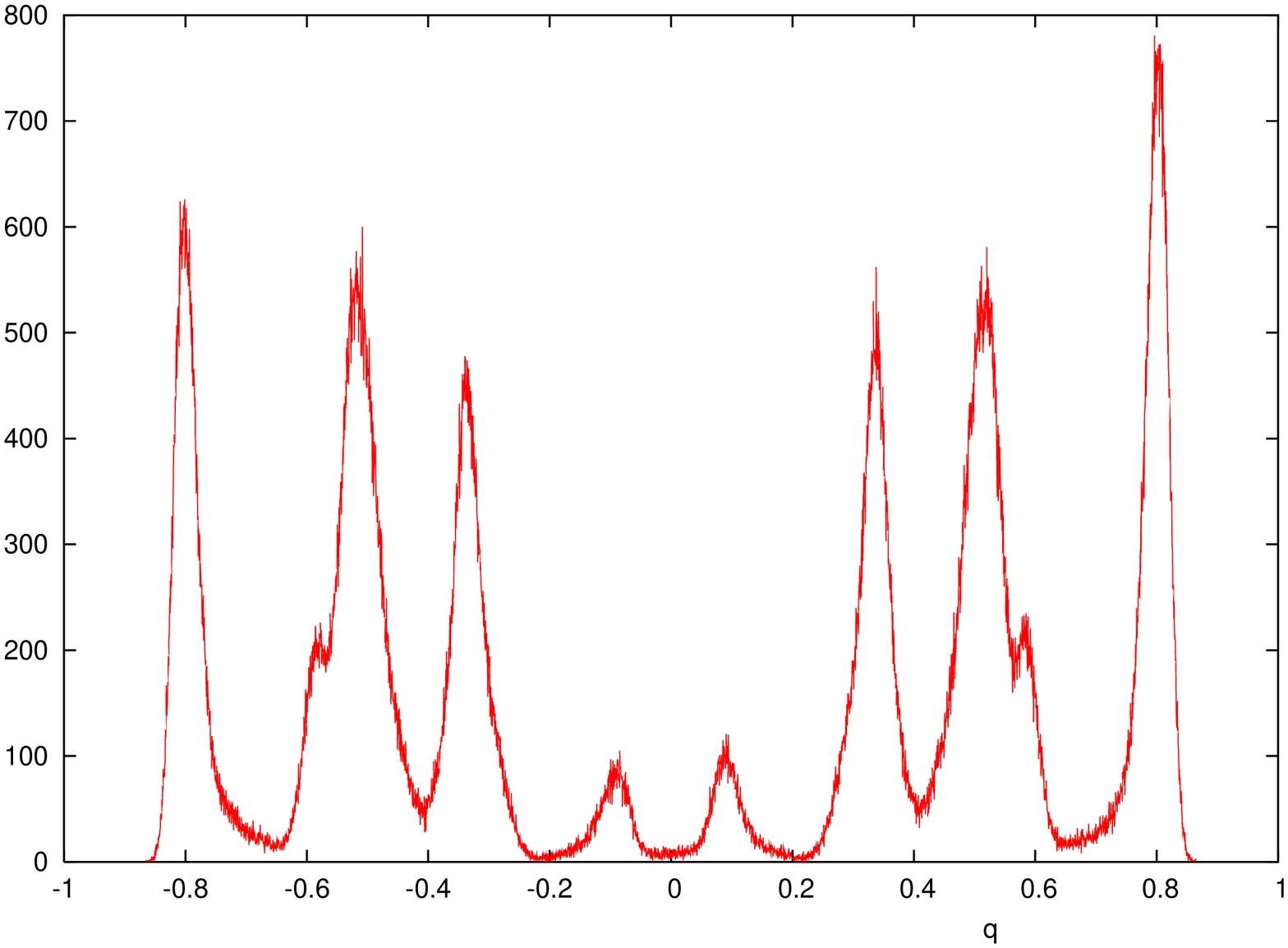}
\hskip 2cm
\includegraphics*[height=4.5cm,angle=270]{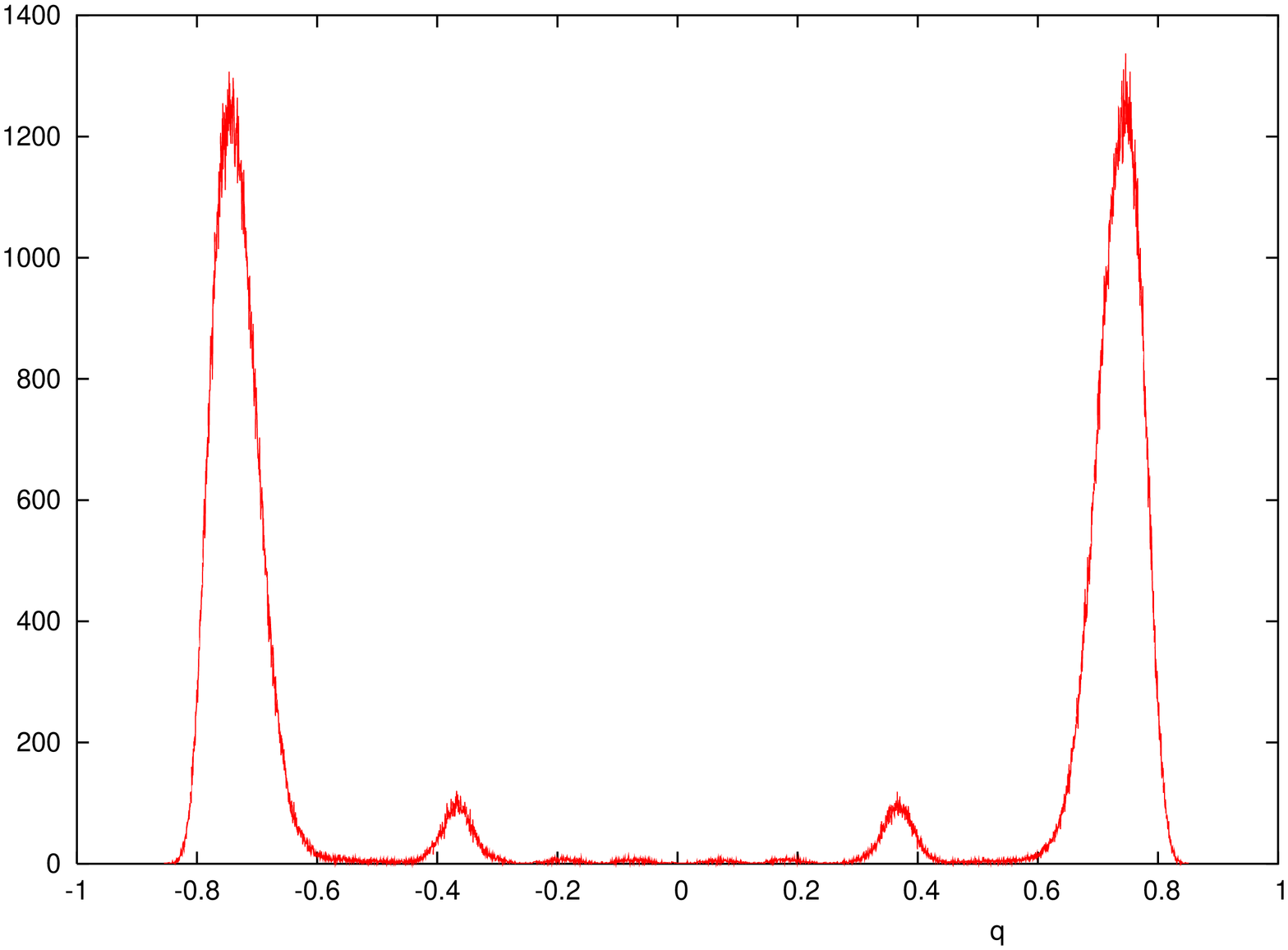}
\caption{$P_J(q)$ for eight different disorder realization. Here $N=4096$ and
$T=0.4$}
\label{fig:3}       
\end{figure}

Figure~\ref{fig:3} shows $P_J(q)$ for eight different disorder
realizations, using the data of~\cite{BiMa}. Here $N=4096$ and
$T=0.4$, a pair of values that is at the borderline of what can be
done with current algorithms and computer resources.  We are well
inside the spin-glass phase, and the number of spins is equal to the
number of spins of a $16^3$ lattice~\footnote{This is to say that it
is a large system, by current spin glass standards.}. These eight
samples are just the eight first samples generated by our computer
program, they have not been selected afterwards.  One sees clearly
that the shape of $P_J(q)$ is strongly fluctuating from sample to
sample. In view of~(\ref{replica7}) this is very suggestive of the
existence of several (disorder dependent) pure states. The two extreme
peaks correspond to the self overlap $q_{EA}$ (which is the same for
all pure states), the other peaks correspond to cross overlaps.  If
the shape of $P_j(q)$ is strongly fluctuating, the position of the
outmost peak is not, in agreement with the prediction that $q_{EA}$ is
self-averaging.  One notes finally that the curves are reasonably
symmetric, this is a very strong sign that the Monte Carlo statistics is
sufficient.

\begin{figure}
\centering
\includegraphics[width=0.414\textwidth,angle=270]{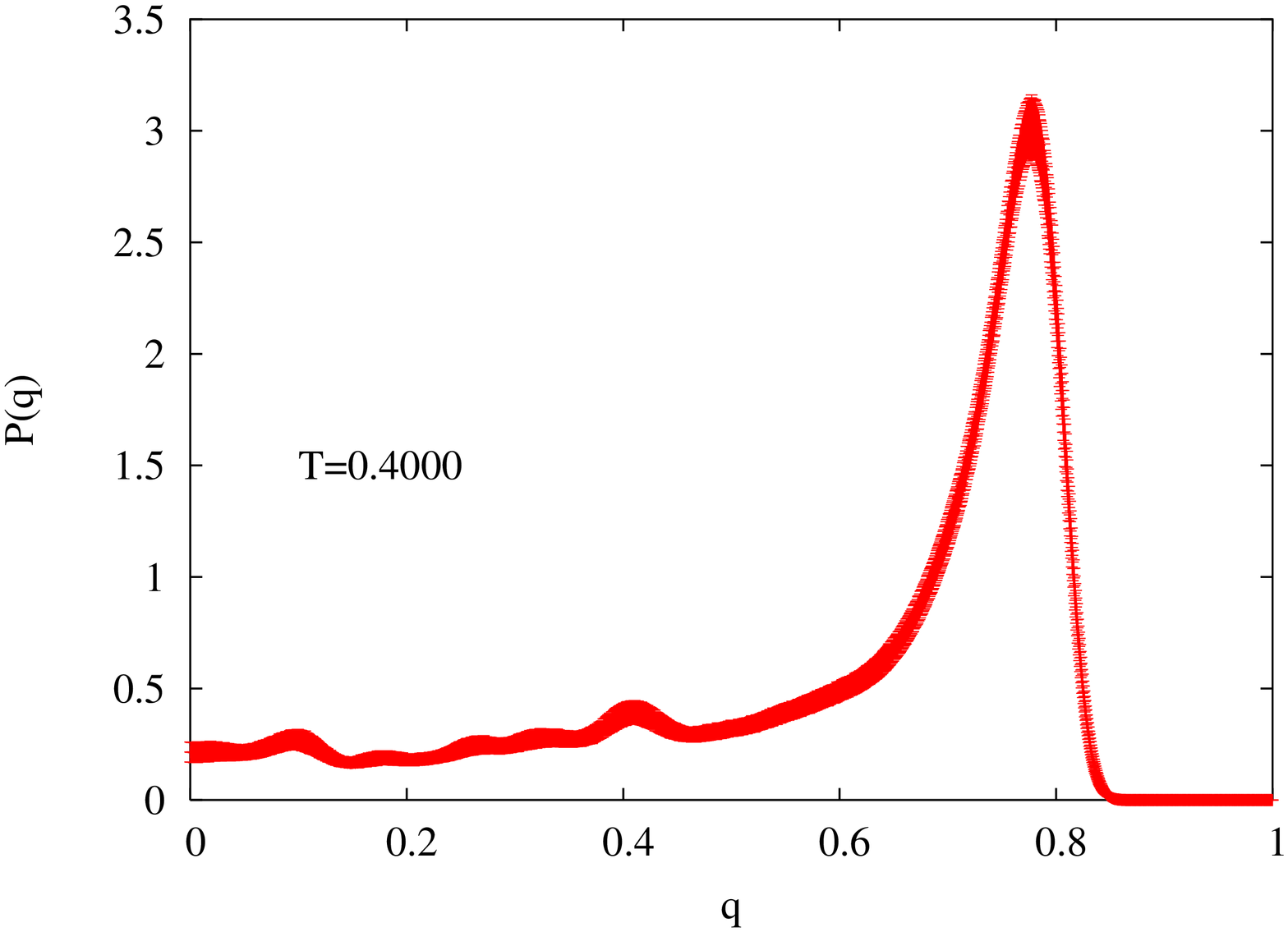}
\caption{The disorder averaged $P(q)$ for $q>0$ and zero magnetic field. Here $N=4096$ and
$T=0.4$. The wiggles are a fluctuation due to the finite (256) number
of disorder samples.  For this value of $T$, the infinite volume limit
of $q_{EA}$ is~\cite{CrRiz2} $q_{EA}=0.759$.}
\label{fig:4}       
\end{figure}

The non self averaging of $P_J(q)$ makes the 
measurement of the
disorder averaged $P(q)$ quite difficult  on large lattices. This is
exemplified in Fig.~\ref{fig:4}, which presents our estimate of this
average with all  $256$ disorder
samples. The wiggles are artifacts due to the finite number of
samples.

\begin{figure}
\centering
\includegraphics[width=0.414\textwidth,angle=270]{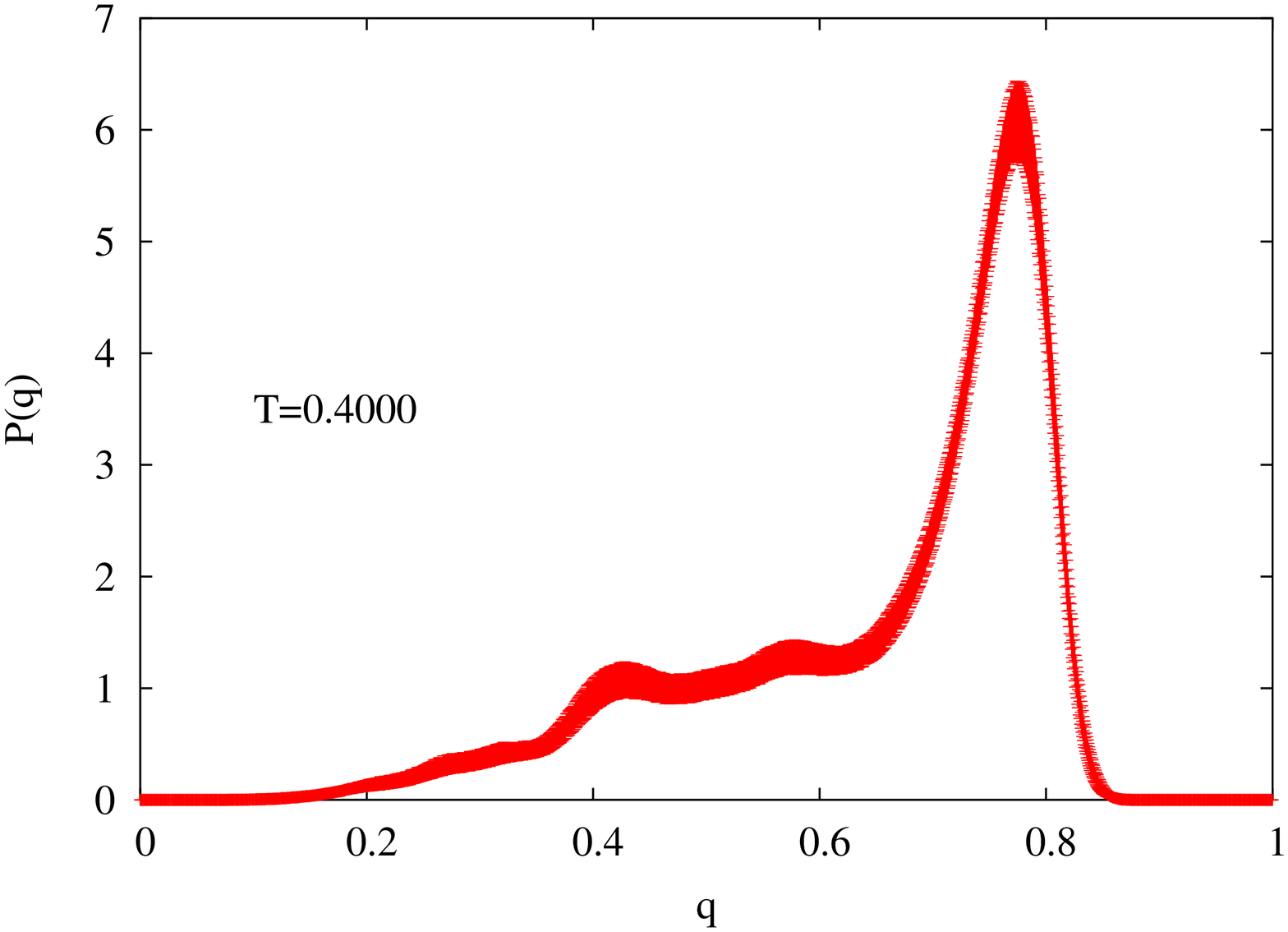}
\caption{The disorder averaged $P(q)$ for $q>0$ at nonzero $H=0.3$. Here $N=3200$ and
$T=0.4$. For these values of $H$ and $T$, the infinite volume limit of
$q_{EA}$ and $q_{min}$ are~\cite{CrRiz2} $q_{EA}=0.759$ and
$q_{min}\approx 0.44$.  The self-overlap $q_{EA}$ is nearly $H$
independent at fixed $T$. This is a consequence of the so called
Parisi--Toulouse hypothesis
.}
\label{fig:5}       
\end{figure}

The Parisi solution has the quite unusual prediction that the low
temperature phase extends to nonzero $H$ (up to the AT line). This
prediction is unfortunately hard to check numerically, as can be seen
in Fig.~\ref{fig:5} from~\cite{BiCo}.  Even the very existence of a
peak at $q_{min}$ is not clear from the data.  In order to see this
peak unambiguously, one would need to satisfy two conflicting
constrains, $H$ should not be too small, since the weight of the peak
goes to zero as $H\to 0$.  It should also not be too close to the $AT$
line since $q_{min} \to q_{max}$ in this limit.  Clearly much larger
systems would bee needed in order to see a clear distinct peak at
$q_{min}$.

A classical method to locate the critical point from numerical data
uses the Binder parameter (the Kurtosis of the order parameter
distribution function). For the SK model this is

\begin{equation}
B_N(H,T)= {1 \over 2} \left ( 3 - {\overline{\langle (q-\overline{\langle q \rangle})^4 
\rangle} \over
\overline{\langle (q-\overline{\langle q \rangle})^2 \rangle}^2} \right )\;.
\label{binder}
\end{equation}
Here $B(H,T)$ is defined~\footnote{When $H=0$ $<q>=0$, and the formula
simplifies substantially.}  in such a way that it is zero if $P_J(q)$
is Gaussian (namely at high enough temperature) and one for a two
equal weight delta functions distribution. In finite dimension we know
from finite size scaling that $B_N(H,T)$ is a function of $(T-T_c(H))
L^{1/\nu}=(T-T_c(H))N^{1/(\nu d)}$, and accordingly the curves for
$B_N(H,T)$ drawn for different values of $N$ should all cross at the critical
point $T_c(H)$.  This provides a very convenient numerical method to
determine a critical point.  In the SK model, finite size scaling
holds with $d=6$ and $\nu=1/2$ (see Subsec.~\ref{subsec:12}). The
Binder parameter method to locate $T_c$ works nicely at zero magnetic
field, the curves for different values of $N$ decrease monotonously as
a function of $T$, are well separated away from $T_c$ and cross nicely
at $T_c$. This is not the case at nonzero magnetic field~\cite{BiCo}, at
least with system sizes one can currently simulate.

It is fair to admit that, from a numerical perspective, the AT line
remains elusive.  This is not a problem in the SK model case since
there is no doubt about the existence and exact location of this
transition line.  It become annoying however if one has the EAI case
in mind.

I will close this section by mentioning that there are some numerical
evidences for ultrametricity, in the sense that triplets of typical
spin configurations (with the same disorder configuration) $\{
\sigma_i^{(1)} \}$,$\{ \sigma_i^{(2)} \}$,$\{ \sigma_i^{(3)} \}$ do fulfill~\cite{Ultra}
the ultrametric constraint.  There is however no numerical evidences~\cite{Tree}
for the full treelike structure of states, as predicted by Parisi solution.

\section{Finite size effects for the free energy and the internal energy}

Numerical simulations are obviously limited to finite systems, and
simulations of spin glasses are indeed limited to very small systems.
A detailed understanding of finite size effects in spin glass models
is accordingly highly desirable.  In what follows, I will only
consider the free energy and the internal energy at zero magnetic
field.  I define the exponents $\mu$ and $\omega$ according to the
equations

\begin{eqnarray}
f_N(T)&=&f(T)+B(T) N^{-\mu}+\cdots\\
\nonumber
e_N(T)&=&e(T)+C(T) N^{-\omega}+\cdots\;,
\end{eqnarray}
where $f(T)$ and $e(T)$ are the infinite volume free energy and
internal energy at temperature $T$.

\subsection{Paramagnetic phase}
\label{subsec:20}

In the high temperature phase, the finite size effects for the
internal energy and free energy  can be obtained from the
truncated model of~(\ref{trunc1},\ref{trunc2}).
At first order, one keeps only the quadratic term in $\myL[\qtilda ]$, with the result

\begin{eqnarray}
\nonumber
\beta f_N(T)
&=&-\ln{2}-\frac{\beta^2}{4}
-\lim_{n\to0}\frac{1}{n N}\ln\int \Biggl[\prod_{a< b}\sqrt{\frac{N}{2\pi\beta^2}}\
\nonumber
d\qtilda_{ab}
\Biggr]\exp(-N \tau \sum_{a< b}q^2_{a,b})\\
&=&-\ln{2}-\frac{\beta^2}{4}
-\lim_{n\to0}\frac{1}{n N}\ln\Biggl[\prod_{a< b}\sqrt{\frac{1}{2\tau\beta^2}}\Biggr]\\
&=&-\ln{2}-\frac{\beta^2}{4}
-\frac{1}{4 N}\ln{{2\tau\beta^2}}\;.
\end{eqnarray}

The neglected terms in $\myL[\qtilda ]$ can be included by expanding
the exponent as a power
series~\cite{PRS1,PRS2,YMA}. Each term is represented by
a graph without external leg. Each line gives a factor $1/(N\tau)$,
and each vertex a factor $N$. Introducing the number of lines $\#L$,
of vertices $\#V$ and loops $\#B$ of a given graph, one finds that the
graph behaves like
$N^{\#V-1}(N\tau)^{-\#L}=N^{-\#B}\tau^{-\#L}$. Organizing the
expansion as a loop expansion, one obtains an expansion in powers of
$1/N$, with the most singular term (as $\tau\to 0$) at each order,
given by the contribution of the cubic term in $\myL$ (for which
$\#L=3(\#B-1)$). The development up to the $O(1/N^4)$ order (four
loops) has been obtained in~\cite{PRS1} (the terms of order $1/N^4$
are not given in the paper, but have been used in the re-summation at
$T_c$). Results up to five loops, for arbitrary $n$ but omitting the
quartic terms in $\myL$ can be found in~\cite{YMA}.

The internal energy is simply obtained through the equation

\begin{equation}
e_N(T)=\frac{d}{d\beta}f_N(T),
\end{equation}
that holds also at finite $N$.

As one approaches the critical point both expansions for $f_N(T)$ and
$e_N(T)$ become singular, and need to be re-summed.  Writing symbolically the $N$ dependent
part of $f_N(T\approx T_c)$ as

\begin{eqnarray}
& &
-\lim_{n\to0}\frac{1}{n N}\ln\int \Biggl[\prod_{a< b}\sqrt{\frac{N}{2\pi}}\
d\qtilda_{ab}
\Biggr]\exp(-N (\tau Q^2+Q^3+ Q^4))\\
\nonumber
&=&
-\lim_{n\to0}\frac{1}{n N}\ln\int \Biggl[\prod_{a< b}\sqrt{\frac{(xN)^{1/3}}{2\pi}}\
d\qtilda_{ab}
\Biggr]\exp(-(Q^2+x^{1/2}Q^3+ x^{2/3}/N^{1/3}Q^4))\;,
\end{eqnarray}
where $x=1/(N\tau^3)$, we obtain the $N$ dependent part of $f_N(T_c)$ as
the $x\to\infty$ limit of

\begin{eqnarray}
\nonumber
& & \frac{\ln N}{12 N} 
-\frac{1}{N}
\lim_{n\to0}\frac{1}{n}\ln\int \Biggl[\prod_{a< b}\sqrt{\frac{x^{1/3}}{2\pi}}\
d\qtilda_{ab}
\Biggr]\exp(-(Q^2+x^{1/2}Q^3+ x^{2/3}/N^{1/3}Q^4))\;.
\end{eqnarray}

Treating the 
order four term as a perturbation one obtain finally

\begin{equation}
\displaystyle {f_N(T_c)}=-\ln 2-1/4+
\frac{\ln N}{12 N}+\frac{f_{(-1)}}{N}+\frac{f_{(-4/3)}}{N^{4/3}} +\cdots \;,
\label{FTC}
\end{equation}

\begin{equation}
e_N(T_c)=-\frac{1}{2}+  \frac{e_{(-2/3)}}{N^{2/3}}+O(N^{-1})\;.
\end{equation}

The behavior of the internal energy is the one expected from
scaling. At a critical point, the singular part of the internal energy
for a $L^d$ system behaves~\footnote{This follows from the scaling
expression $f=1/L^d \tilde{F}((T-T_c)L^{1/\nu})$.}  like
$L^{1/\nu-d}$. Using the values $\nu=1/2$ and $d=6$ (see
Subsec.~\ref{subsec:12}) one finds that the singular part of the  internal energy does
behave like $N^{-2/3}$.  In order to obtain the values of the
coefficients $f_{(-1)}$, and $e_{(-2/3)}$, one needs to handle the
theory in the strong coupling $x\to\infty$ regime. This is not easy
and theses values are poorly determined~\cite{PRS1}.

\subsection{Zero temperature}

With the discovery of very efficient algorithms for determining the
ground state of a spin glass finite system (at fixed $\myJ$), the
physics of zero temperature spin glass has blossomed in the recent
years.  This includes detailed studies of the finite size effects of
the internal energy.  In the specific case of the SK model, good
evidences have been obtained for a $1/N^{2/3}$ behavior for the
internal energy for both Gaussian and binary disorder
distributions~\cite{Palassini1,Palassini2, BKM, Boettcher,KKLJH,Pal}, as
summarized in Tab.~\ref{table}.  The value $\omega=2/3$ is exact for
the spherical Sherrington--Kirkpatrick model~\cite{AnBaMa}.

\begin{table}[htb]
\begin{center}
\begin{tabular}{|l|l|r|r|}
\hline
 			&   $P(\myJ)$	& $N_{max}$ &	$\omega$	   		\\
\hline
Palassini~\cite{Palassini2}  	& Gaussian 	& $199$ 	& $0.673 \pm 0.002$	\\
Bouchaud et al~\cite{BKM}	& $\pm 1$	& $300$		& $0.66 \pm 0.02$ 	\\
Boettcher~\cite{Boettcher}	& $\pm 1$	& $1023$	& $0.672 \pm 0.005$ 	\\  
Katzgraber et al~\cite{KKLJH}	& Gaussian	& $192$		& $0.64 \pm 0.01$	\\
Pal~\cite{Pal}			& both		& $2048$	& $\approx 2/3$		\\
\hline
\end{tabular}\end{center}
\caption {Numerical estimates for the exponent $\omega$ at $T=0$ for the Sherrington--Kirkpatrick model: 
reference, disorder distribution, maximum system size and result.}
\label{table}
\end{table}

\subsection{Using the Guerra and  Toninelli formalism}

As shown in~\cite{moi}, one can use the so called Guerra and Toninelli
interpolation formalism, an ingredient in the proof~\cite{GuTo} that
the free density energy of the Sherrington--Kirkpatrick model
converges in the infinite volume limit N, as the basis of a powerful
numerical method to compute the finite size corrections to the free
energy of the model.  Guerra and Toninelli introduced the partition
function

\begin{eqnarray}
Z_N(t)&=&\sum_{\{\sigma\}}\exp\Biggl({\beta}\Bigl(\sqrt{\frac tN}\sum_{1\le i<j\le N}
J_{ij}\sigma_i\sigma_j\\
\nonumber
&+&\sqrt{\frac {1-t}{N/2}}\sum_{1\le i<j\le N/2}
J'_{ij}\sigma_i\sigma_j\\\nonumber
&+&\sqrt{\frac {1-t}{N/2}}\sum_{N/2 <i<j\le N}
J''_{ij}\sigma_i\sigma_j\Bigr)\Biggr)
\exp(\beta H \sum_i \sigma_i)\;,
\end{eqnarray}
that involves a parameter $t$ that interpolates between the SK model
with $N$ sites ($t=1$) and a system of two un-coupled SK models with
$N/2$ sites ($t=0$).  The $J_{i,j}$'s, $J'_{i,j}$'s and $J''_{i,j}$'s
are independent identically distributed Gaussian random numbers.  It
is straightforward to show that

\begin{eqnarray}
\beta(f_{N}(T)-f_{N/2}(T))&=&\frac{\beta^2}{4}\int_0^1 dt 
\overline{
\Med{(q_{12})^2-\frac{1}{2}(q_{12}^{(1)})^2-
\frac{1}{2}(q_{12}^{(2)})^2}}
\label{DeltaF}
\\
\nonumber
&=&\frac{\beta^2}{4}\int_0^1 dt\ {\myD}(t)\qquad \myD(t) \leq 0,
\end{eqnarray}
where

\begin{eqnarray}
q_{12}&=&\frac{1}{N} \sum_{i=1}^N \sigma_i\tau_i, \quad
q_{12}^{(1)}=\frac{2}{N} \sum_{i=1}^{N/2} \sigma_i\tau_i\;, \\
\nonumber
q_{12}^{(2)}&=&\frac{2}{N} \sum_{i=N/2+1}^{N} \sigma_i\tau_i\;.
\end{eqnarray}

In the above formulas, the $\sigma_i$'s and $\tau_i$'s are the spins
of two real replica, $q_{12}$ is the usual overlap, for the $N$ sites
system, $q_{12}^{(1)}$ and $q_{12}^{(2)}$ are the overlaps restricted
to the two subsystems with $N/2$ sites.  The right hand side
of~(\ref{DeltaF}) can be evaluated with Monte Carlo simulation.  I use
the Parallel Tempering algorithm, with $T\in [0.4,1.3]$ and uniform
$\Delta T=0.025$.  A total of $2\ 10^5$ sweeps of the algorithm was
used for every disorder sample.  The quenched couplings have a binary
distribution in order to speed up the computer program. Using the
cumulant expansion~(\ref{cumulant}) one can show that in the
paramagnetic phase, the replacement of Gaussian couplings by binary
couplings induces a leading $1/N$ correction to the free energy
density, of the same order as the leading finite size correction (see
below). One can argue that the effect of the binary couplings is also
$1/N$ in the spin glass phase, and is thus negligible with respect to
the leading $1/N^{2/3}$ finite size correction: In the
spin glass phase the leading finite size correction is the same for
the binary and Gaussian couplings.  Systems of sizes $N$ from $128$ to
$1024$ have been simulated with $128$ disorder samples for each system
size (but for $N=1024$, where I used $196$ samples).  The integration
over $t$ was done with the trapezoidal rule, with $39$ non uniformly
spaced points~\footnote{With a distribution adapted to the shape of
$\myD(t)$, that is peaked at low $t$.}. Integrating with only half of
the points makes a very small effect on the integrand (smaller than
the estimated statistical error).  The data presented at $T_c$
(Fig.~\ref{figure4} and~\ref{figure6}) include the results of an
additional simulation of a system with $N=2048$ sites, limited to the
paramagnetic phase, with $T\in [1.0,1.3]$, $\Delta T=0.025$, with 128
disorder samples, and a $15$ points discretization of $t$.

\begin{figure}[tbh]
\centering
\includegraphics[width=0.414\textwidth,angle=270]{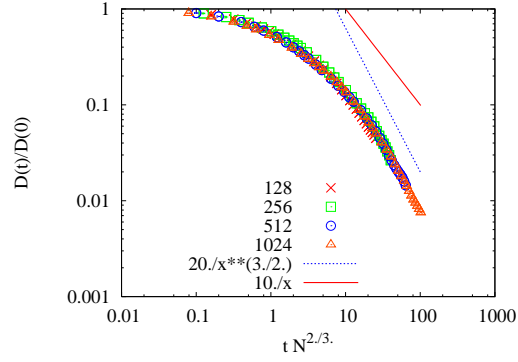}
  \caption{ ${\myD(t)}/{D(t=0)}$ as a function of $t
  N^{2/3}$ (both in logarithmic scale), for $T=0.6$.  The 
  full line shows the $1/x$ behavior, the dotted line
  shows the $1/x^{3/2}$ behavior. Clearly $D(t)$ decreases faster than
  $1/x$ for large $x$. The precise behavior of $\myD(t)$ is not
  essential for my argument, as soon as $\myD(t)$ decays faster than $1/x$.}
\label{figure2}
\end{figure}

\begin{figure}[tbh]
  \centering
  \includegraphics[width=0.414\textwidth,angle=270]{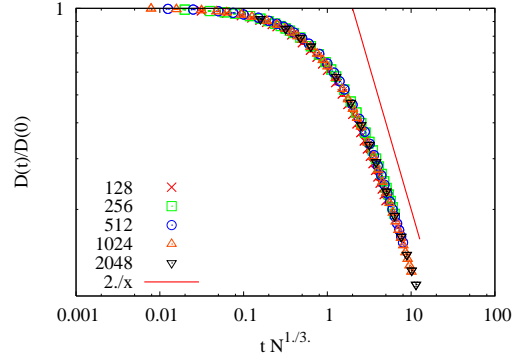}
  \caption{ ${\myD(t)}/{D(t=0)}$ as a function of $t
  N^{1/3}$ (both in logarithmic scale), for $T=T_c$.  The 
  straight line shows the expected $1/x$ behavior, in order to guide
  the eyes.}
\label{figure4}
\end{figure}

In the low $T$ phase, a remarkable scaling is observed if one plots
the ratio ${\myD(t)}/{D(t=0)}$ as a function of $t N^{2/3}$, as shown in
Fig.~\ref{figure2}.  It means that, to a good approximation, one
has $\myD(t)/\myD(0)=F(t N^{2/3})$, with the function $F(x)$ decaying faster than $1/x$
for large $x$, making the integral in~(\ref{DeltaF}) converge.
One has accordingly in the low $T$ phase $f_N-f_{\infty}\propto
1/N^{2/3}$.  

The situation is different at $T_c$, as shown in Fig.~\ref{figure4},
the ratio ${\myD(t)}/{D(t=0)}$ scales with a different exponent,
namely like $F(t N^{1/3})$, with a large $x$ behavior
compatible~\footnote{Admittedly much larger system sizes would be
needed in order to be sure that the system really obeys this
asymptotic behavior.} with $F(x)\propto 1/x$, then 

\begin{eqnarray}
f_{N}(T)-f_{N/2}(T)&\propto&\int_0^1 dt\ {\myD}(t)
={\myD}(0)\int_0^1 dt\ F(tN^{1/3})\\
\nonumber
&\propto& 1/N \ \ln{N/N_0}\;,
\end{eqnarray}
for some undetermined $N_0$. Use has been made of the fact that,
according to finite size scaling, $\myD(0)=
-1/2\overline{<(q_{1,2}^{(1)})^2>} =
-1/2\overline{<(q_{1,2}^{(2)})^2>} $ scales like
$N^{2\beta/(d\nu)}\tilde{G}((T-T_c)N^{1/(d\nu)})$, with a finite non
zero $\lim_{x\to0}\tilde{G}(x)$ and that $\beta/(d\nu)=2/3$ (see
Subsec.~\ref{subsec:12}).  This behavior of $f_{N}(T)-f_{N/2}(T)$ is
in agreement\footnote{But the value of the prefactor is not
predicted by my method.} with~(\ref{FTC}).

\begin{figure}[tbh]
  \centering
  \includegraphics[width=0.414\textwidth,angle=270]{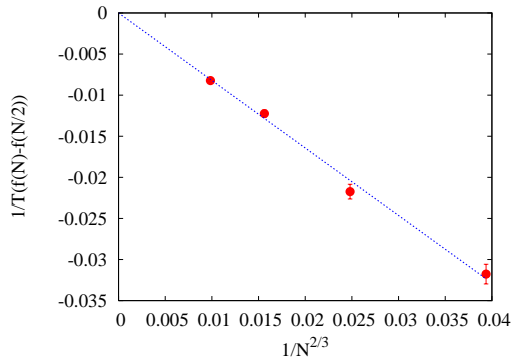}
  \caption{ Numerical data for $(f_N-f_{N/2})/T$ as a
  function of $1/N^{2/3}$, together with a numerical fit to the data of
  the form $(f_N-f_{N/2})/T=-A/N^{2/3}$, with $A=0.82\pm 0.02$ 
  dotted line.  Here $T=0.4$, $N=128, 256, 512\ {\rm and}\ 1024$.}
\label{figure5}
\end{figure}

Figure~\ref{figure5} shows, as a function of $1/N^{2/3}$, my estimate
at $T=0.4$ of the difference $(f_N-f_{N/2})/T$, obtained by
integrating numerically~(\ref{DeltaF}), compared to the result of a
linear fit $(f_N-f_{N/2})/T= - A/N^{-2/3}$, with $A= 0.82\pm 0.02$ and
$\chi^2=4.9$.  The agreement is good within estimated statistical
errors.  A similar agreement is obtained for other values of $T$ in
the spin glass phase, e.g.  $A=0.39 \pm 0.01$ with $\chi^2=3.6$ for
$T=0.6$, and $A=0.18 \pm 0.01$ with $\chi^2=33$ (a large value
presumably related to the proximity of the critical point) for
$T=0.8$.  The value $\omega=2/3$ is in contradiction with the old
analytical prediction of~\cite{CrPaSovu}.  In this paper, arguments
are given that relate $\omega$ to the exponent $p$ of the first
correction term in the expansion of the replicated free energy
$-1/\beta\lim_{N\to\infty}1/N\ln\overline {Z_{\myJ}^n}$ in powers of
$n$, by the equation $\omega=(p-1)/p$. In the paramagnetic phase it is
known that there is no term in this expansion beyond the linear term,
i.e. $p$ is infinite, and thus $\omega=1$ in agreement with the
previous discussion.  In the spin glass phase Kondor~\cite{Kondor} has
found, using the truncated model of~(\ref{trunc1}), that $p=6$, and
thus $\omega=5/6$. The resolution of this contradiction lays
presumably in the use of the Parisi approximation~(\ref{trunc3})
in~\cite{Kondor}, and we can conjecture from our data that indeed
$p=3$.

\begin{figure}[tbh]
  \centering
  \includegraphics[width=0.414\textwidth,angle=270]{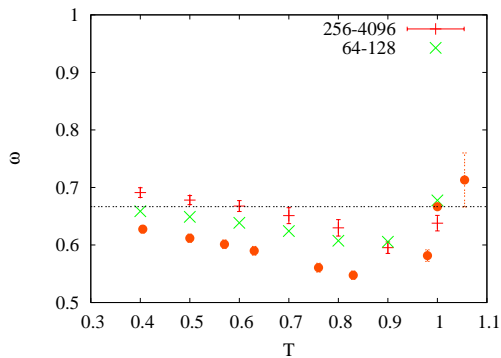}
  \caption{Behavior of the finite size exponent $\omega$ as a function
  of temperature for the Sherrington--Kirkpatrick model. From top to
  bottom: results of a (gnuplot) fit of the data of~\cite{BiMa} with
  $4096 \geq N \geq 256$ (with estimated errors); fit of the same data
  with $128\geq N \geq 64$ (without estimated errors, in order not to
  clutter the figure) and data from~\cite{KaCa}. }
\label{figure7}
\end{figure}

\begin{figure}[tbh]
  \centering
  \includegraphics[width=0.414\textwidth,angle=270]{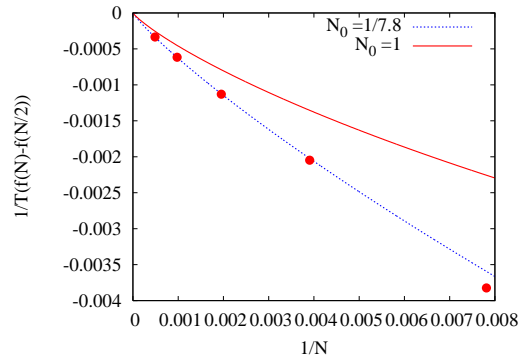}
  \caption{ Numerical data for $(f_N-f_{N/2})/T$ as a
  function of $1/N$, together with the behavior implied by the
  equation: $f_N/T=f_{\infty}/T+1/(12 N)\ln{N/N_0}$. The 
  full line is drawn with the value $N_0=1$. The dotted
  line is drawn with the value $1/7.8$, from a fit to the data.  Here
  $T=1$, $N=128, 256, \ldots,2048$.}
\label{figure6}
\end{figure}

Since the energy at zero temperature (and thus the free energy) also
behaves like $1/N^{2/3}$, the most natural conclusion is that the
leading finite size corrections for both $f_N(T)$ and $e_N(T)$ behave
like $1/N^{2/3}$ in the whole low temperature phase, for both binary
and Gaussian distributions. This seems however to contradict the
numerical results of~\cite{KaCa}. In this paper, the internal energy
of the SK model with Gaussian $P(\myJ)$ has been measured with Monte
Carlo simulations for values of $T$ between $T=0.1$ (a very low value
made possible by the small sizes simulated) and $T=1.22$, with $N=36,
64, \ldots, 196$. The data for $e_N(T)$ are fitted~\footnote{Using the
precise estimates of $e(T)$ from~\cite{CrRiz}.} as
$e_N(T)=e(T)+a(T)N^{-\omega}$, with a result for $\omega$ that is
compatible with $2/3$ for both $T=0$ and $T=T_c$, but with a
pronounced deep between, in contradiction with my conjecture. Using
the data for $e_N(T)$ obtained from the simulation of~\cite{BiMa}, it
is quite simple to repeat the analysis of~\cite{KaCa} for systems up
to $N=4096$ (but with binary $P(\myJ)$).  Figure~\ref{figure7} shows
my results compared to the one of~\cite{KaCa}.  The use of larger
system sizes clearly reduce the effect observed in this paper.  The
most likely conclusion is that the analysis of~\cite{KaCa} is affected
by systematic errors due to the small sizes used. The shape of the
sub-leading corrections to the energy is not known below $T_c$, it is
known at the critical point, however, and there the sub-leading
corrections are decaying very slowly, the expansion is indeed an
expansion in powers of $N^{-1/3}$, and it is may be not so surprising
to have difficulties finding the correct leading exponent from data
with $36\leq N \leq 196$.

Figure~\ref{figure6} shows my estimates for $(f_N-f_{N/2})/T$ at $T_c$
as a function of $1/N$, together with the prediction of~(\ref{FTC}).
A good agreement (with $\chi^2=4.3$ if one excludes the $N=128$ data
from the fit) is obtained using the value $1/N_0=7.8\pm 0.2$, namely
$f_{(-1)}=\ln(7.8)/12=0.17\ldots$, within the range of results
presented in~\cite{PRS1}.

The method has been extended~\cite{GA} to the computation of the
fluctuations of the free energy, with the result that, in the spin
glass phase, $\Delta^2 f_N(T)=\overline {f^2_N(T)}-\overline
{f_N(T)}^2\propto N^{-2\sigma}$, with $\sigma\approx 3/5$. This is
definitively smaller than the value found at $T=0$~\cite{Palassini2,
BKM, Boettcher,KKLJH,Pal}, that is $\sigma\approx 3/4$.

\subsection{Conclusions}

The picture that emerges is the following: above $T_c$ the finite size
corrections of both $f_N(T)$ and $e_N(T)$ are given by series in
powers of $1/N$, whose terms can be evaluated by perturbation theory.
This expansion becomes singular as the critical point is
approached. At the critical point the size dependent part of the free
energy behaves like $(1/12N)\ln{N/N_0}$ and the size dependent part of
the internal energy as $e_{(-2/3)}/N^{2/3}$, with coefficients that
are poorly determined analytically, but can be evaluated with Monte
Carlo simulation. The sub-leading corrections are decaying very slowly
with $N$.  In the spin glass phase, the data are compatible with a
$1/N^{2/3}$ behavior for both the leading finite size corrections to
$e_N(T)$ and $f_N(T)$.

\section{Acknowledgments}

I thank Wolfhard Janke for organizing this very pleasant CECAM
workshop.  My interest in the Sherrington--Kirkpatrick model stems
from a continued collaboration with Enzo Marinari, which I thanks
warmly. The $H\neq 0$ data have been obtained in collaboration with
Barbara Coluzzi.  I also thank Andrea Crisanti and Tommaso Rizzo and
Helmut Katzgraber, and Ian Campbell for providing me with their
numerical data, and Giulio Biroli, Jean-Philippe Bouchaud, Cirano de Dominicis and Tam\`as Temesv\'ari for discussions  . I am deeply indebted to Giorgio Parisi for many
insights.

\end{document}